\documentclass[aps,prd,11pt,superscriptaddress,notitlepage,longbibliography,nofootinbib,tightenlines,preprintnumbers]{revtex4-1}
\usepackage{amsmath,amssymb,amsfonts}
\usepackage{graphicx}
\usepackage{color}
\usepackage{tikz}
\usepackage{booktabs}
\usepackage{dsfont}
\usepackage[pdftex]{hyperref}
\hypersetup{colorlinks=true, linkcolor=darkred, citecolor=blue, linktoc=page}
\definecolor{darkred}{rgb}{0.8,0.1,0.1}

\usetikzlibrary{decorations.markings}

\usepackage{subfigure}

\def\cA{{\cal A}}
\def\cB{{\cal B}}

\def\cD{{\cal D}}

\def\cG{{\cal G}}
\def\cI{{\cal I}}
\def\cJ{{\cal J}}

\def\RR{\ensuremath{\mathbb R}}

\def\cL{{\cal L}}

\def\cZ{{\cal Z}}

\def\cL{{\cal L}}
\def\cN{{\cal N}}

\def\CC{{\mathds{C}}}

\DeclareMathOperator{\vol}{vol}

\DeclareMathOperator{\Li}{Li}

\makeatletter\def\l@subsubsection#1#2{}%
\makeatother

\newcommand{\nocontentsline}[3]{}
\newcommand{\tocless}[2]{\bgroup\let\addcontentsline=\nocontentsline#1{#2}\egroup}

\def\Re{\mathop{\rm Re}}

\begin{document}

\title{Evidence for a 5d F-theorem}

\author{Martin Fluder}
\email{mfluder@princeton.edu}
\affiliation{Joseph Henry Laboratories\\ Princeton University, Princeton, NJ 08544, USA\\[2mm]}

\author{Christoph F.~Uhlemann} 
\email{uhlemann@umich.edu}

\affiliation{Leinweber Center for Theoretical Physics, Department of Physics
	\\
	University of Michigan, Ann Arbor, MI 48109-1040, USA}

\preprint{LCTP-20-25}

\begin{abstract}
 Renormalization group flows are studied between 5d SCFTs engineered by $(p,q)$ 5-brane webs with large numbers of external 5-branes. A general expression for the free energy on $S^5$ in terms of single-valued trilogarithm functions is derived from their supergravity duals, which are characterized by the 5-brane charges and additional geometric parameters. The additional geometric parameters are fixed by regularity conditions, and we show that the solutions to the regularity conditions extremize a trial free energy. These results are used to survey a large sample of $\mathcal O(10^5)$ renormalization group flows between different 5d SCFTs, including Higgs branch flows and flows that preserve the $SU(2)$ $R$-symmetry. In all cases the free energy changes monotonically towards the infrared, in line with a 5d $F$-theorem.
\end{abstract}

\maketitle

\tableofcontents

\clearpage

\setlength{\parskip}{1.5 pt}

\section{Introduction}

Conformal field theories (CFTs) are important landmarks in the space of general quantum field theories (QFTs) and are studied extensively. 
A natural next step in charting out the landscape of QFTs more generally is to understand renormalization group (RG) flows connecting different CFTs.
RG flows are harder to analyze, due to the lack of conformal symmetry in between the end points, which makes general results particularly valuable. 
A striking class of such general results are ``c-theorems'', which identify quantities that change monotonically along RG flows.
Such ``c-functions'' strongly constrain the possible flows between different CFTs, indicate the irreversibility of RG flows, and they give useful measures for the number of degrees of freedom of a given system, which is hard to define otherwise in QFT. Furthermore, in certain limits c-functions can be directly related to the density of states, \emph{e.g.} by Cardy formulae~\cite{Cardy:1986ie}, and they feature prominently in the context of AdS/CFT and black hole microstate counting.

Monotonicity properties for different types of c-functions have been established in $d=2,3,4$~\cite{Zamolodchikov:1986gt,Cardy:1988cwa,Komargodski:2011vj,Jafferis:2010un,Jafferis:2011zi,Klebanov:2011gs,Casini:2012ei},
and for certain types of flows in supersymmetric theories in $d=6$~\cite{Maxfield:2012aw,Elvang:2012st,Heckman:2015axa,Cordova:2015vwa,Cordova:2015fha}. 
The remaining dimension in which supersymmetric conformal field theories (SCFTs) exist, $d=5$, is much less understood.
A holographic c-theorem holds in general dimensions for flows described by certain truncated gravity theories~\cite{Myers:2010xs,Myers:2010tj}, but while highly indicative, the flows that can actually be realized within consistent truncations to these gravity theories are rather limited.

In this work we revisit RG flows between 5d SCFTs.
Many 5d SCFTs can be understood as UV fixed points of (perturbatively non-renormalizable) 5d gauge theories \cite{Seiberg:1996bd,Intriligator:1997pq,Bhardwaj:2020gyu}, and even larger classes of theories can be realized in string theory (recent classification attempts include \cite{Jefferson:2017ahm,Jefferson:2018irk,Bhardwaj:2018yhy,Bhardwaj:2018vuu,Apruzzi:2018nre,Apruzzi:2019vpe,Apruzzi:2019opn,Apruzzi:2019enx,Bhardwaj:2019xeg,Apruzzi:2019kgb}).
We will focus on a large general class of 5d SCFTs that can be engineered by $(p,q)$ 5-brane webs in Type IIB string theory~\cite{Aharony:1997ju,Aharony:1997bh}.
Akin to 3d~\cite{Jafferis:2010un,Jafferis:2011zi,Klebanov:2011gs,Casini:2012ei}, the quantity that is expected to be monotonic along RG flows is the free energy on the 5-sphere~\cite{Klebanov:2011gs},
\begin{align}\label{eqn:Fdef}
F  = - \log \cZ_{S^{5}} \,.
\end{align}
Consequently, results establishing the monotonicity of $F$ are also referred to as $F$-theorems.
Certain flows between 5d SCFTs have been found to be compatible with a putative $F$-theorem at finite $N$ in~\cite{Chang:2017cdx}, and at large $N$ \emph{e.g.}\ in~\cite{Jafferis:2012iv,Fluder:2018chf,Fluder:2019szh,Uhlemann:2019ypp}. However, unlike in 3d no general proof for the monotonicity of $F$ is available,\footnote{The methods based on entanglement entropy used to prove $F$/$a$-theorems in 3d/4d in \cite{Casini:2012ei,Casini:2017vbe} do not extend straightforwardly to 5d, since the relation between $F$ and the sphere entanglement entropy includes a third order derivative of the entanglement entropy whose positivity properties are harder to access.} 
and the data supporting the conjectured monotonicity is somewhat sparse.

Several properties are desirable for a function to be identified as a measure for the number of degrees of freedom: 
$(i)$ it should be bounded from below 
$(ii)$ it should be invariant under marginal deformations
$(iii)$ it should decrease along RG flows from the UV to the IR.
The last property can be formulated in various forms, the weakest of which is that the value at the IR fixed point should be smaller than the value at the UV fixed point. 
Concretely in 5d, the putative c-function $-F$ is expected to satisfy
\begin{align}\label{eqn:Fthm}
	-F_{\rm UV} > - F_{\rm IR}~.
\end{align} 
Property $(i)$ is in general not satisfied by $-F$ in 5d:
free Maxwell theory, which is not conformal in 5d \cite{ElShowk:2011gz}, has $-F = -\infty$ \cite{Giombi:2015haa}, showing that $-F$ is unbounded from below at least if non-conformal theories are not excluded.
Marginal deformations are scarce in 5d (there are no supersymmetric ones), but some evidence for property $(ii)$ has been given in \cite{Klebanov:2011gs}.
In this paper we focus on (\ref{eqn:Fthm}), and provide extensive evidence that it holds for RG flows between large classes of large-$N$ 5d SCFTs.\footnote{In 3d, statements analogous to (i) and (iii) have been proven. In fact, a stronger version of (\ref{eqn:Fthm}) holds: $F$ is defined and monotonic also in between the fixed points. While there is evidence for property (ii), a proof is lacking.}

Our approach is to study RG flows between general 5d SCFTs engineered by $(p,q)$ 5-brane webs with large numbers of external branes in Type IIB string theory, using the holographic duals for these theories constructed in~\cite{DHoker:2016ujz,DHoker:2016ysh,DHoker:2017mds}.
Using a combination of analytic and numerical methods, we survey a large sample of RG flows.
The free energies $-F$ obtained from the supergravity duals are manifestly positive, and have been matched to field theory computations for numerous examples in \cite{Fluder:2018chf,Uhlemann:2019ypp}.
We compare the free energies between the UV and IR fixed points of the RG flows, and find that, for all flows considered, (\ref{eqn:Fthm}) is satisfied, in line with the existence of a 5d $F$-theorem.

We will derive a general expression for the sphere free energy of 5d SCFTs engineered by $(p,q)$ 5-brane webs with large numbers of external 5-branes from their supergravity duals. 
This free energy is expressed in terms of the charges of the $(p,q)$ 5-branes involved in the string theory realization of the 5d SCFTs, and additional parameters characterizing the geometry of the internal space in the supergravity duals. These additional parameters are fixed in terms of the 5-brane  charges by regularity conditions. If the regularity conditions are not satisfied, the supergravity duals are singular, and there is no immediate field theory interpretation. 
Nevertheless, our expression for the free energy with the geometric parameters left unconstrained is finite and can be regarded as a ``trial free energy''.
We show that solutions to the regularity conditions extremize this trial free energy.
Using these results, we show analytically that the free energy decreases for certain simple classes of RG flows, and survey a large sample of more general flows numerically.

\medskip

\textbf{Outline:} In Section~\ref{sec:5dscfts}, we review relevant aspects of 5d SCFTs engineered by $(p,q)$ 5-brane webs, and discuss RG flows from that perspective. In Section~\ref{sec:sugra-F}, we review the planar limit of 5d SCFTs and summarize the main results on the free energies, whose derivation we relegate to appendices. In Section~\ref{sec:flows}, we discuss explicit RG flows. 
We close with a discussion in Section~\ref{sec:disc}. 


\section{5d SCFTs and RG flows}\label{sec:5dscfts}

\begin{figure}
	\subfigure[][]{\label{fig:5-brane-gen}
		\begin{tikzpicture}[scale=1.5]
			
			\foreach \i in {-1,0,1}{\foreach \j in {100,-50}{
					\draw[very thick] ({sin(\j)*\i*0.1},{-cos(\j)*\i*0.1}) -- ({cos(\j)+sin(\j)*\i*0.1},{sin(\j)-cos(\j)*\i*0.1}); 
			}}
			
			\foreach \i in {-3/4,3/4}{\foreach \j in {30,-110,-190}{
					\draw[very thick] ({sin(\j)*\i*0.1},{-cos(\j)*\i*0.1}) -- ({cos(\j)+sin(\j)*\i*0.1},{sin(\j)-cos(\j)*\i*0.1}); 
			}}
			
			\draw[fill=gray] (0,0) circle (5pt);
			\node[anchor=south] at ({cos(100)},{sin(100)}) {\scriptsize $(p_1,q_1)$};
			\node [anchor=south west] at ({cos(30)},{sin(30)})  {\scriptsize  $(p_2,q_2)$};
			\node [anchor=north west] at ({cos(-50)},{sin(-50)}) {\scriptsize  $(p_3,q_3)$};	
			\node [anchor=east] at ({cos(-190)},{sin(-190)}) {\scriptsize  $(p_L,q_L)$};	
			\node [rotate=125] at (-0.9,-0.5) {\ldots};
			\draw[->] (-1.3,-1) -- (-1.3+.25,-1) node [below,font=\footnotesize] {x};
			\draw[->] (-1.3,-1) -- (-1.3,-1+.25) node [left,font=\footnotesize] {y};
		\end{tikzpicture}
	}
	\qquad\qquad
	\subfigure[][]{\label{fig:plus-web}
		\begin{tikzpicture}[scale=0.8]
			\foreach \i in {-1.2,-0.6,0,0.6,1.2}{
				\draw[thick] (\i,-1.8) -- (\i,1.8);}
				\foreach \j in {-0.75,-0.25,0.25,0.75}{
					\draw[thick] (2.2,\j) -- (-2.2,\j);
				}
				
			\node at (0,-2.2) {\small $M$\,NS5};
			\node at (-2.9,0) {\small $N$\,D5};
			\draw[->] (-2.5,-2.5) -- (-2.5+.5,-2.5) node [below,font=\footnotesize] {x};
			\draw[->] (-2.5,-2.5) -- (-2.5,-2.5+.5) node [left,font=\footnotesize] {y};
		\end{tikzpicture}
	}
	\caption{Left: Generic 5-brane junction, fully specified by a choice of external 5-branes. The external branes have been resolved slightly for illustrative purposes; the figure represents a junction at a point. The branes are at angles reflecting their charge, \emph{i.e.}\ $\Delta x/\Delta y=p/q$.
		Right: Mass deformation of a junction of $M$ NS5 and $N$ D5-branes.
		The UV fixed point is described by the configuration with all D5 and all NS5 branes coincident (see Figure~\ref{fig:flavor-mass}). For $N=M=1$ this describes a free massless hypermultiplet.
		\label{fig:junction}}
\end{figure}
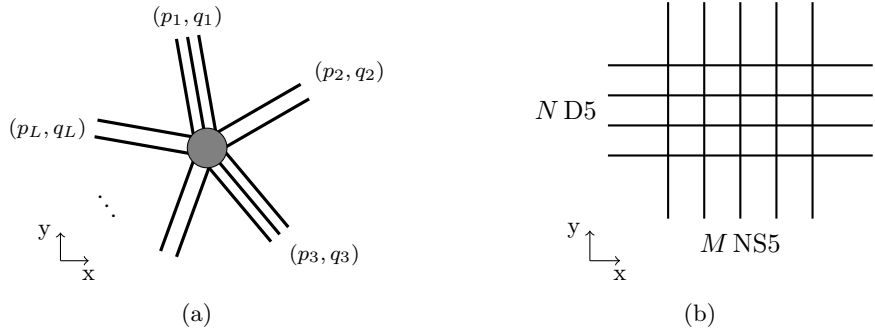

Many 5d SCFTs can be realized as UV fixed points of perturbatively non-renormalizable 5d gauge theories with $\cN=1$ supersymmetry.\footnote{The superconformal algebra in 5d is unique and includes 8 Poincar\'e supercharges, which fixes the amount of supersymmetry for theories with UV fixed points in 5d.}
We will start here with a string theory construction of 5d SCFTs in terms of 5-brane webs, which also covers SCFTs that can not be described as UV fixed points of gauge theories, and make connections to gauge theory descriptions along the way. In this section we review the features of 5-brane webs that will be pertinent below; more details can be found for instance in~\cite{Aharony:1997ju,Aharony:1997bh}.

The basic ingredients are $(p,q)$ 5-branes of Type IIB string theory, where $(1,0)$ is a D5 and $(0,1)$ an NS5-brane. Supersymmetric configurations of $(p,q)$ 5-branes can be realized if all branes share five of their six worldvolume dimensions, and have the remaining one lying in a common plane, say with coordinates $x$ and $y$, at an angle dictated by the $(p,q)$ charges, \emph{i.e.}~$p\Delta x=q\Delta y$. These 5-branes can join and split so long as their $(p,q)$-charges are conserved.
Every planar junction of such $(p,q)$ 5-branes at a point in the $xy$-plane defines a 5d SCFT with 16 supersymmetries (comprising 8 Poincar\'e supersymmetries and their superconformal partners).
An example is shown schematically in Figure~\ref{fig:5-brane-gen}, where the external 5-branes have been resolved slightly for illustrative purposes.
The SCFTs have an $SU(2)$ R-symmetry, which in the string theory realization corresponds to rotations in the three directions transverse to all 5-branes.
Relevant deformations of the SCFT correspond to displacements of a subset of the semi-infinite external 5-branes.

Many 5-brane junctions at a point can be deformed to brane webs involving stacks of parallel D5-branes of finite extent in the horizontal direction, suspended between more general $(p,q)$ 5-branes. 
An example for an intersection of $N$ D5-branes and $M$ NS5-branes, discussed first in \cite{Aharony:1997bh}, is shown in Figure~\ref{fig:plus-web}.
Following \cite{Bergman:2018hin}, we refer to this theory as $+_{N,M}$.
Each stack of $N$ D5-branes corresponds to an $SU(N)$ gauge node, while the semi-infinite D5-branes at the ends realize fundamental hypermultiplets.
A gauge theory for the $+_{N,M}$ junction thus is a linear quiver gauge theory of $SU(N)$ nodes, with neighboring nodes connected by hypermultplets in the bifundamental representation, and with flavors in the fundamental representation at the boundary nodes:\footnote{Another gauge theory description can be obtained by performing an S-duality on the brane web, corresponding to a 90 degree rotation. This leads, in this case, to a gauge theory of the same form but with $M$ and $N$ exchanged.}
\begin{equation}\label{eq:plus-quiver}
[N]-SU(N)-\ldots -SU(N)-[N] \,,
\end{equation}
with a total of $M-1$ gauge nodes and with all Chern-Simons levels zero. In general, a 5-brane junction may or may not have deformations that can be described as 5d gauge theories. However, whether or not a 5d SCFT has deformations described by gauge theories will not play a role in the following. We will only use that relevant deformations are realized by displacements of some of the external 5-branes, which allows for a simple geometric picture of RG flows for general SCFTs.

The 5d SCFTs considered in this paper are fully specified by a choice of 5-brane charges $(p_1,q_1), \ldots, (p_L,q_L)$ defining a 5-brane junction at a point.
These theories have holographic duals based on the Type IIB string theory solutions constructed in~\cite{DHoker:2016ysh,DHoker:2017mds}.\footnote{More general theories can be realized by brane webs involving for example  7-branes and orientifold planes \cite{DeWolfe:1999hj,Bergman:2015dpa,Zafrir:2015ftn}. Supergravity duals for certain classes of such theories are available \cite{DHoker:2017zwj,Uhlemann:2019lge}, but will not be considered here.} Below, we will consider theories for which the supergravity approximation to Type IIB string theory becomes accurate, which is when all (integer) charges $(p_\ell,q_\ell)$ are homogeneously large.
We will generally refer to the limit where all $(p_\ell,q_\ell)$ scale with $N$, \emph{i.e.} are $\mathcal O(N)$ with $N$ large, as the ``large N limit".

For 5-branes with large $(p,q)$ one may distinguish two cases: 1) if $(p,q)$ have a large common divisor, we have a stack of a large number of like-charged 5-branes and 2) if $(p,q)$ are relatively prime, it is a single 5-brane with large charges. The two cases differ substantially in terms of relevant deformations of the SCFTs and also for instance regarding the spectrum of stringy states realizing SCFT operators (studied from the holographic perspective in~\cite{Bergman:2018hin}). However, from the perspective of the free energies, the two cases in general differ only by an $\mathcal O(1)$ adjustment of the brane charges, and the free energies in the large $N$ limit are insensitive to such $\mathcal O(1)$ adjustments. Consequently, we can treat the two cases on equal footing.

In the following we discuss two classes of RG flows from the brane web perspective, to set the stage for the discussion of the change in free energy along such flows in the following sections.

\begin{figure}
\subfigure[][]{\label{fig:HB-flows-1}
	\begin{tikzpicture}[x={(-0.866cm,0.5cm)}, y={(0.866cm,0.5cm)}, z={(0cm,1cm)}, scale=0.9]
		\foreach \i in {-1.2,-0.6,0,0.6,1.2}{
			\draw[thick] (\i,-1.8,0) -- (\i,1.8,0);}
		\foreach \j in {-0.75,-0.25,0.75}{
			\draw[thick] (2.2,\j,0) -- (-2.2,\j,0);
		}
		\draw[thick,blue] (2.2,0.25,1.8) -- (-2.2,0.25,1.8);
		\foreach \i in {-1.2,-0.6,0,0.6,1.2}{	\draw[dashed,blue] (\i,0.25,0) -- (\i,0.25,1.8);}
		
		\draw[->] (-2,-2) -- (-2+.5,-2) node [below,font=\footnotesize] {y};
		\draw[->] (-2,-2) -- (-2,-2+.5) node [below,font=\footnotesize] {x};
		\draw[->] (-2,-2) -- (-2+.5,-2+.5) node [right,font=\footnotesize] {z};
	\end{tikzpicture}
}
\qquad\qquad
\subfigure[][]{\label{fig:HB-flows-2}
	\begin{tikzpicture}[x={(-0.866cm,0.5cm)}, y={(0.866cm,0.5cm)}, z={(0cm,1cm)}, scale=0.9]
	\foreach \i in {-1.2,-0.6,0,1.2}{
		\draw[thick] (\i,-1.8,0) -- (\i,1.8,0);}
	\foreach \j in {-0.75,-0.25,0.75}{
		\draw[thick] (2.2,\j,0) -- (-2.2,\j,0);
	}
	\draw[thick,blue] (2.2,0.25,1.8) -- (-2.2,0.25,1.8);
	\draw[thick,blue] (0.6,-1.8,1.8) -- (0.6,1.8,1.8);
	
	\foreach \i in {-1.2,-0.6,0,1.2}{	\draw[dashed,blue] (\i,0.25,0) -- (\i,0.25,1.8);}
	\foreach \i in {-0.75,-0.25,0.75}{	\draw[dashed,blue] (0.6,\i,0) -- (0.6,\i,1.8);}
	\draw[->] (-2,-2) -- (-2+.5,-2) node [below,font=\footnotesize] {y};
	\draw[->] (-2,-2) -- (-2,-2+.5) node [below,font=\footnotesize] {x};
	\draw[->] (-2,-2) -- (-2+.5,-2+.5) node [right,font=\footnotesize] {z};
	\end{tikzpicture}
}
	\caption{Left: Higgs branch deformation of the $+_{4,5}$ web in Figure~\ref{fig:plus-web} in which a complete D5-brane is moved orthogonally from the plane of the web. The IR fixed point of the flow triggered by this deformation is obtained by moving the D5 off to infinity, leaving a $+_{3,5}$ web. Right: A junction of a D5 and an NS5 is separated from the plane in which the $+_{5,4}$ web of Figure~\ref{fig:plus-web} lies, leaving behind a $+_{3,4}$ web.
		\label{fig:HB-flows}}
\end{figure}
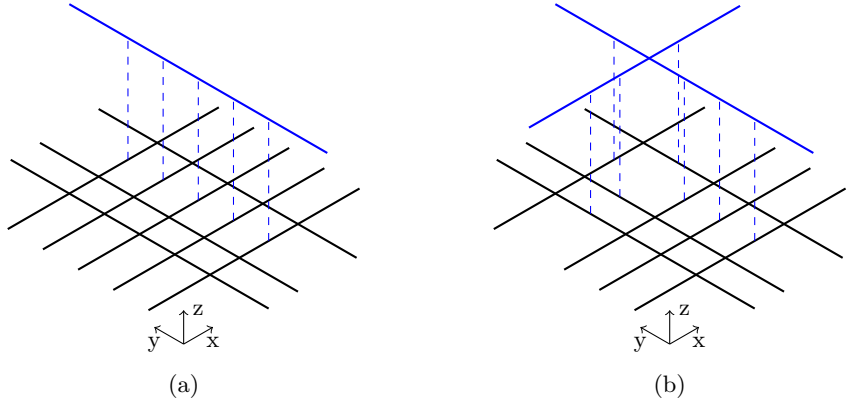

\subsection{Higgs branch flows}

The first class are Higgs branch flows, which are triggered by relevant deformations that break the  $R$-symmetry $SU(2)_R$. The $SU(2)_R$ symmetry corresponds to rotations in the directions transverse to all 5-branes, and Higgs branch deformations are realized by moving a sub-junction of 5-branes out of the $xy$-plane in which the full 5-brane junction lies. 

The simplest case is to separate an entire $(p,q)$ 5-brane from the 5-brane junction: For instance, for the $+_{M,N}$ theory shown in Figure~\ref{fig:plus-web}, one may move an entire D5-brane out of the $xy$-plane. This triggers a flow whose fixed point in the infrared is obtained by moving the 5-brane off to infinity, leaving behind the original junction with one complete 5-brane removed. This is illustrated in Figure~\ref{fig:HB-flows-1} for a flow connecting the $+_{4,5}$ to the $+_{3,5}$ theory.\footnote{A related flow is discussed in Section~IIF of \cite{Aharony:1997bh}. In general, Higgs branch flows are possible only from certain points on the moduli space. Here we only discuss flows between 5d SCFTs at the origins of their moduli spaces.}
Such flows are only possible if the original 5-brane junction involves semi-infinite $(p,q)$ 5-branes joining from opposite directions.

More generally, brane webs which are composed of two independent sub-webs were called reducible in \cite{Aharony:1997bh}. For the 5-brane junctions realizing 5d SCFTs at the origin of their moduli space, reducibility amounts to having subsets of 5-branes whose charges individually sum to zero.
From such a reducible web one may remove any sub-junction of semi-infinite 5-branes whose charges sum to zero.
The natural next case to consider is removing a triple-junction of semi-infinite 5-branes and one may go up to removing a sub-junction of as many 5-branes as the original junction involves.
In general the removed sub-junction describes additional light degrees of freedom and the IR theory comprises two decoupled sectors.
An example where a junction of four semi-infinite external 5-branes is removed is shown in Figure~\ref{fig:HB-flows-2}, depicting a Higgs branch flow connecting the $+_{4,5}$ to a theory consisting of two decoupled sectors, one described by the $+_{3,4}$ theory and one by  the $+_{1,1}$ theory corresponding to a free hypermultiplet.\footnote{Further Higgs branch flows can be realized by terminating the external $(p,q)$ 5-branes on $[p,q]$ 7-branes and moving segments of 5-branes out of the $xy$ plane, as discussed \emph{e.g.} in \cite{Benini:2009gi}. We will not consider such flows here.}

\begin{figure}
	\begin{tikzpicture}[scale=1.4]
		
		\foreach \i in {-1,0,1}{\foreach \j in {0,90,180,270}{
				\draw[very thick] ({sin(\j)*\i*0.1},{-cos(\j)*\i*0.1}) -- ({cos(\j)+sin(\j)*\i*0.1},{sin(\j)-cos(\j)*\i*0.1}); 
		}}
		
		\draw[fill=gray] (0,0) circle (5pt);
	\end{tikzpicture}
	\qquad\qquad
	\begin{tikzpicture}[scale=1.4]
		
		\foreach \i in {-1,0,1}{\foreach \j in {180,270}{
				\draw[very thick] ({sin(\j)*\i*0.1},{-cos(\j)*\i*0.1}) -- ({cos(\j)+sin(\j)*\i*0.1},{sin(\j)-cos(\j)*\i*0.1}); 
		}}
		\foreach \i in {-1,0}{\foreach \j in {90}{
				\draw[very thick] ({sin(\j)*\i*0.1},{-cos(\j)*\i*0.1}) -- ({cos(\j)+sin(\j)*\i*0.1},{sin(\j)-cos(\j)*\i*0.1}); 
		}}
		
		\foreach \i in {0,1}{\foreach \j in {0}{
				\draw[very thick] ({sin(\j)*\i*0.1},{-cos(\j)*\i*0.1}) -- ({cos(\j)+sin(\j)*\i*0.1},{sin(\j)-cos(\j)*\i*0.1}); 
		}}
		
		\draw[very thick] (0,0) -- ({0.7*cos(45)},{0.7*sin(45)}) -- (1,{0.7*sin(45)}) ; 
		\draw[very thick] ({0.7*cos(45)},{0.7*sin(45)}) -- ({0.7*cos(45)},1) ; 
		
		\draw[fill=gray] (0,0) circle (5pt);
	\end{tikzpicture}
	\qquad\qquad
	\begin{tikzpicture}[scale=1.4]
		\foreach \i in {-1,0,1}{\foreach \j in {180,270}{
				\draw[very thick] ({sin(\j)*\i*0.1},{-cos(\j)*\i*0.1}) -- ({cos(\j)+sin(\j)*\i*0.1},{sin(\j)-cos(\j)*\i*0.1}); 
		}}
		\foreach \i in {-1,0}{\foreach \j in {90}{
				\draw[very thick] ({sin(\j)*\i*0.1},{-cos(\j)*\i*0.1}) -- ({cos(\j)+sin(\j)*\i*0.1},{sin(\j)-cos(\j)*\i*0.1}); 
		}}
		
		\foreach \i in {0,1}{\foreach \j in {0}{
				\draw[very thick] ({sin(\j)*\i*0.1},{-cos(\j)*\i*0.1}) -- ({cos(\j)+sin(\j)*\i*0.1},{sin(\j)-cos(\j)*\i*0.1}); 
		}}
		
		\draw[very thick] (0,0) -- ({cos(45)},{sin(45)});
		
		\draw[fill=gray] (0,0) circle (5pt);
	\end{tikzpicture}
	
	\caption{Left: $+_{N,M}$ junction. Middle: A D5-brane from is joined with an NS5 brane and separated from the junction, creating a  $(1,1)$ 5-brane of finite extent. In the gauge theory (\ref{eq:plus-quiver}) this deformation corresponds to a flavor mass term. Right: Moving the D5/NS5 combination off to infinity decouples the flavor hypermultiplet.\label{fig:flavor-mass}}
\end{figure}
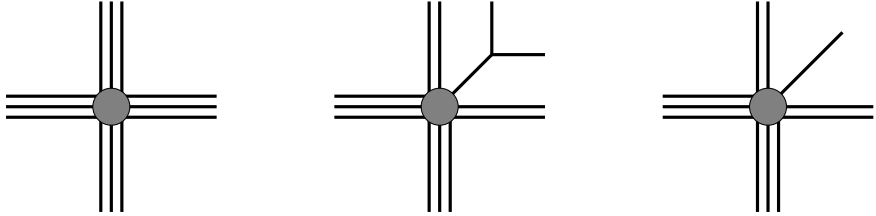

\subsection{Mass deformations}\label{sec:mass-flow}

\begin{figure}
	\begin{tikzpicture}
\begin{scope}[rotate=-90,scale=0.6]
		\draw[thick] (-0.25,-2.5) -- (-0.25,1.5) -- (0.25,1.5) -- (0.25,-2.5);
		\draw[thick] (-0.25,1.5) -- +(-1.25,1.25);
		\draw[thick] (0.25,1.5) -- +(1.25,1.25);
		\draw[thick] (-0.75,-2.5) -- (-0.75,0.75) -- (0.75,0.75) -- (0.75,-2.5);
		\draw[thick] (-0.75,0.75) -- +(-1.25,1.25);
		\draw[thick] (0.75,0.75) -- +(1.25,1.25);
		\draw[thick] (-1.25,-2.5) -- (-1.25,0) -- (1.25,0) -- (1.25,-2.5);
		\draw[thick] (-1.25,0) -- +(-1.25,1.25);
		\draw[thick] (1.25,0) -- +(1.25,1.25);
		\draw[thick] (-1.75,-2.5) -- (-1.75,-0.75) -- (1.75,-0.75) -- (1.75,-2.5);
		\draw[thick] (-1.75,-0.75) -- +(-1.25,1.25);
		\draw[thick] (1.75,-0.75) -- +(1.25,1.25);
\end{scope}
	\end{tikzpicture}
\qquad\qquad
	\begin{tikzpicture}
	\begin{scope}[rotate=-90,scale=0.6]
		\draw[thick] (-0.25,-2.5) -- (-0.25,2);
		\draw[thick] (0.25,-2.5) -- (0.25,2);
		\draw[thick] (-0.75,-2.5) -- (-0.75,0.75) -- (0.75,0.75) -- (0.75,-2.5);
		\draw[thick] (-0.75,0.75) -- +(-1.25,1.25);
		\draw[thick] (0.75,0.75) -- +(1.25,1.25);
		\draw[thick] (-1.25,-2.5) -- (-1.25,0) -- (1.25,0) -- (1.25,-2.5);
		\draw[thick] (-1.25,0) -- +(-1.25,1.25);
		\draw[thick] (1.25,0) -- +(1.25,1.25);
		\draw[thick] (-1.75,-2.5) -- (-1.75,-0.75) -- (1.75,-0.75) -- (1.75,-2.5);
		\draw[thick] (-1.75,-0.75) -- +(-1.25,1.25);
		\draw[thick] (1.75,-0.75) -- +(1.25,1.25);
		
		\draw[dashed] (-0.25,2) -- (-0.25,3) -- (0.25,3);
		\draw[dashed] (0.25,2) -- (0.25,3);
		\draw[dashed] (-0.25,3) -- +(-1.25,1.25);
		\draw[dashed] (0.25,3) -- +(1.25,1.25);
		
	\end{scope}
\end{tikzpicture}
\qquad\qquad
\begin{tikzpicture}
	\begin{scope}[rotate=-90,scale=0.6]
	\draw[thick] (-0.25,-2.5) -- (-0.25,2);
	\draw[thick] (0.25,-2.5) -- (0.25,2);
	\draw[thick] (-0.75,-2.5) -- (-0.75,2);
	\draw[thick] (0.75,-2.5) -- (0.75,2);	
	\draw[thick] (-1.25,-2.5) -- (-1.25,0) -- (1.25,0) -- (1.25,-2.5);
	\draw[thick] (-1.25,0) -- +(-1.25,1.25);
	\draw[thick] (1.25,0) -- +(1.25,1.25);
	\draw[thick] (-1.75,-2.5) -- (-1.75,-0.75) -- (1.75,-0.75) -- (1.75,-2.5);
	\draw[thick] (-1.75,-0.75) -- +(-1.25,1.25);
	\draw[thick] (1.75,-0.75) -- +(1.25,1.25);
	
	\draw[dashed] (-0.25,2) -- (-0.25,3.5) -- (0.25,3.5);
	\draw[dashed] (0.25,2) -- (0.25,3.5);
	\draw[dashed] (-0.25,3.5) -- +(-1.25,1.25);
	\draw[dashed] (0.25,3.5) -- +(1.25,1.25);
	
	\draw[dashed] (-0.75,2) -- (-0.75,2.75) -- (0.75,2.75) -- (0.75,2);
	\draw[dashed] (-0.75,2.75) -- +(-1.25,1.25);
	\draw[dashed] (0.75,2.75) -- +(1.25,1.25);
	
	\end{scope}
\end{tikzpicture}

\caption{Relevant deformations of a triple junction involving $(1,0)$, $(1,1)$ and $(1,-1)$ 5-branes (left), leading to quadruple junctions (center and right) upon moving the dashed segments off to infinity. 
	The gauge theory on the left is $[8]-SU(6)-SU(4)-SU(2)$. 
	The web in the center is obtained by running the $SU(2)$ gauge coupling to zero.
The web on the right is obtained by running the $SU(4)$ gauge coupling to zero; the gauge theory described by the remaining solid part is $[8]-SU(6)-[4]$,
the dashed part describes $[4]-SU(2)$.
\label{fig:YN-def}
}
\end{figure}
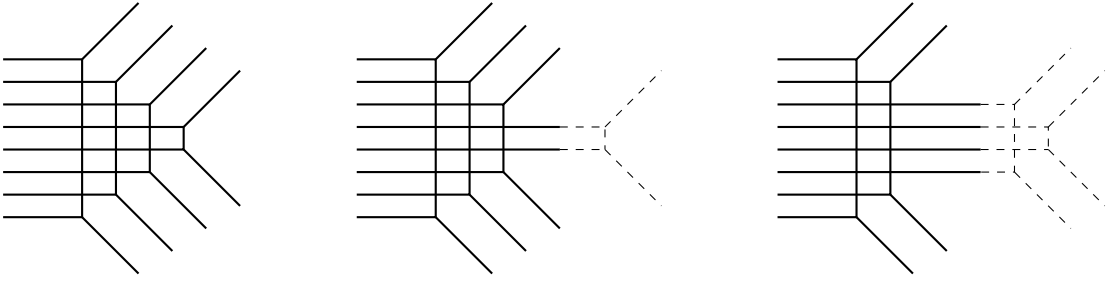

The second class of flows we are interested in are relevant deformations that preserve the $SU(2)_R$ symmetry. 
Examples in gauge theory language include turning on flavor mass terms (see Figure~\ref{fig:flavor-mass}) and turning on a finite (as opposed to infinite) gauge coupling (see Figure~\ref{fig:YN-def}).
We will simply refer to general $SU(2)_R$-preserving relevant deformations as mass deformations.
These deformations can be realized from the brane web perspective by moving a number $K$ of branes in one stack, say of charge $(p,q)$, off to infinity.
For this to be compatible with charge conservation, one may pick a neighboring stack, of charge $(p^\prime,q^\prime)$, and move the $K$ $(p,q)$ 5-branes off along this stack, by merging the $K$ $(p,q)$ 5-branes with $K^\prime$ of the $(p^\prime,q^\prime)$ 5-branes. 
This creates a number $\hat K$ of new branes, with charges $(\hat p,\hat q)$.
The result at the end point of the flow is that $K$ $(p,q)$ 5-branes are eliminated, $K^\prime$ $(p^\prime,q^\prime)$ 5-branes are eliminated, and $\hat K$ $(\hat p,\hat q)$ 5-branes are produced in the original junction, in such a way that
\begin{align}
	K(p,q)+K^\prime(p^\prime,q^\prime)&=\hat K (\hat p,\hat q)~.
\end{align}
This equality follows from charge conservation.
The modified original junction is coupled via an infinitely long link to a junction of $K$ $(p,q)$ 5-branes, $K^\prime$ $(p^\prime,q^\prime)$-branes and $\hat K$ $(-\hat p,-\hat q)$ 5-branes, which in general describes additional 5d degrees of freedom.
The constraints for realizing a relevant deformation are that $K$, $K^\prime$, $\hat K$ are all positive. Charge quantization also constrains $K$, $K^\prime$ and $(\hat p,\hat q)$ to be integers.

Examples are shown in Figure~\ref{fig:flavor-mass} and \ref{fig:YN-def}.
For the flow in Figure~\ref{fig:flavor-mass}, starting from the $+_{N,M}$ junction, semi-infinite 5-branes with $(p,q)=(1,0)$ and $(p^\prime,q^\prime)=(0,1)$ are moved off to infinity, in such a way that  $(\hat p,\hat q)=(1,1)$ 5-branes are created. Charge conservation amounts to $K=K^\prime=\hat K$.
For $K=1$ the relevant deformation corresponds to a mass term for one of the fundamental flavors attached to the last node in the gauge theory (\ref{eq:plus-quiver}).
For $K>1$ the interpretation is more involved.
There are no light 5d degrees of freedom associated with the triple junction that is moved off to infinity for $K=1$.
Figure~\ref{fig:YN-def} on the left shows a triple junction involving $2N$ D5-branes, $N$ $(1,1)$ and $N$ $(1,-1)$ 5-branes, for $N=4$. 
For general $N$ the SCFT was dubbed $Y_N$ theory in~\cite{Bergman:2018hin}. It is the UV fixed point of the quiver gauge theory
\begin{align}
	[2N]-SU(2N-2)-SU(2N-4)-\ldots -SU(2)~.
\end{align}
The deformation shown in Figure~\ref{fig:YN-def} corresponds to joining
a subset of $(p,q)=(1,1)$ and $(p^\prime,q^\prime)=(1,-1)$ 5-branes, to form $(\hat p,\hat q)=(1,0)$ 5-branes. Charge conservation requires $K+K^\prime=2\hat K$.
For general $K$, the relevant deformation amounts to turning on a finite (as opposed to infinite) gauge coupling for the $SU(2K)$ gauge node.
The IR theory resulting from this flow consists of two sectors. 
One sector (the dashed part in Figure~\ref{fig:YN-def}) is a $Y_{K}$ theory, the other is the UV fixed point of the gauge theory
\begin{align}
	[2N]-SU(2N-2)-\ldots - SU(2K+2)-[2K]~.
\end{align}
Both sectors have an $SU(2K)$ symmetry which is weakly gauged, corresponding to the long link between the two webs.

\section{Free energy from supergravity duals}\label{sec:sugra-F}

We now discuss the supergravity duals for 5d SCFTs engineered by 5-brane junctions and their free energies.
In Section~\ref{sec:duals}, the relevant aspects of the solutions are reviewed.
The main results for the free energies are summarized in Section~\ref{eq:F-extr}, while their derivations are relegated to appendices \ref{sec:F-dev} and \ref{app:F-extr}.

\subsection{Supergravity duals}\label{sec:duals}

The geometry of the supergravity duals takes the form of a warped product of $AdS_6$ and $S^2$ over a Riemann surface $\Sigma$, which for the solutions considered here is a disk or, equivalently, the upper half plane, $\mathbb{H}$, without punctures.\footnote{The representations on the disk and on the upper half plane are related by a M\"obius transformation.}
The supergravity duals are defined in terms of a pair of locally holomorphic functions $\cA_\pm (w)$, with $w$ a complex coordinate on $\Sigma = \mathbb{H}$~\cite{DHoker:2016ysh,DHoker:2016ujz,DHoker:2017mds}. 
The differentials, $\partial\cA_{\pm}$, have poles at locations $r_\ell$ on the real line, with residues given by $Z_\pm^\ell$.
Each pole represents an external $(p_\ell,q_\ell)$ 5-brane of the associated brane junction.
The functions  $\cA_\pm (w)$ are given in terms of the pole positions $r_{\ell}$ and the respective residues $Z_\pm^\ell$ by
\begin{align}\label{eqn:cA}
	\cA_\pm (w) &=\cA_\pm^0+\sum_{\ell=1}^L Z_\pm^\ell \ln(w-r_\ell)~,
	&
	Z_\pm^\ell&=\frac{3}{4}\alpha^\prime \left(\pm q_\ell+ ip_\ell\right)~.
\end{align}
The pole positions $r_\ell$ and constants $\cA_\pm^0$ of a given solution are determined in terms of the residues by regularity conditions.
For a solution with $L$ poles these include $L$ conditions, which, for $k=1,\ldots, L$, take the form
\begin{align}	\label{eqn:constr}
	\cA_+^0 Z_-^k - \cA_-^0 Z_+^k 
	+ \sum _{\stackrel{\ell=1}{\ell \neq k} }^LZ^{[\ell, k]} \ln |r_\ell - r_k| &=0~,
	&Z^{[\ell,k]}&\equiv Z_+^\ell Z_-^k - Z_+^k Z_-^\ell~.
\end{align}
The sum over these equations vanishes, since $\sum_\ell Z^{[\ell,k]}=\sum_\ell Z_\pm^\ell=0$ by charge conservation,
leaving $L-1$ independent conditions.
One may solve two of them for the constants $\cA_\pm^0$; the remaining $L-3$ regularity conditions are spelled out in (\ref{eq:constr-4}).
Three poles can thus be fixed arbitrarily, reflecting the $SL(2,\RR)$ automorphisms of the upper half plane, and the remaining ones are fixed by the remaining $L-3$ regularity conditions.
Importantly, there is an additional constraint, which requires that $\partial_w\cA_-$ is non-zero in the interior of $\Sigma$, \emph{i.e.} in the upper half plane,
\begin{align}\label{eq:reg-zeros}
	\partial_w\cA_-&\neq 0~, \qquad \forall w\in {\rm int}(\Sigma)~.
\end{align}
In other words, all zeros of the meromorphic differential $\partial_w\cA_-$ have to be in the lower half plane or on the real line.
We refer to \cite{DHoker:2017mds} for a detailed discussion.
In the supergravity solutions this guarantees that $\kappa^2$ is positive in the interior of $\Sigma$. In terms of the brane picture it enforces that the ordering of the poles, $r_{\ell}$, along $\partial\Sigma$ is compatible with the ordering of the 5-branes forming a particular junction.
Together, the constraints (\ref{eqn:constr}) and (\ref{eq:reg-zeros}) fix all parameters in $\cA_\pm$ in terms of the 5-brane charges, in line with a one-to-one correspondence between 5-brane junctions and supergravity solutions.
From the field theory perspective the absence of moduli reflects that 5d SCFTs have no supersymmetric exactly marginal deformations \cite{Cordova:2016emh}.

The non-trivial Type IIB supergravity fields are the metric $g_{\mu\nu}$, the axio-dilaton $\tau$ and the two-form field $C_{(2)}$, which are given in terms of the functions $\cA_\pm$ as follows
\begin{align}\label{eqn:ansatz}
	ds^2 &= \sqrt{6\cG T} \, ds^2 _{\mathrm{AdS}_6} + \frac{1}{9}\sqrt{6\cG}\,T ^{-\tfrac{3}{2}} \, ds^2 _{\mathrm{S}^2} 
	+ \frac{4\kappa^2}{\sqrt{6\cG}} T^{\tfrac{1}{2}}\, |dw|^2~,
	\nonumber\\
 \tau&=i\,\frac{R\partial_w\cG \partial_{\bar w}(\bar\cA_++\bar\cA_-)-\partial_{\bar w}\cG\partial_w(\cA_++\cA_-)}{R\partial_w\cG\partial_{\bar w}(\bar\cA_+-\bar\cA_-)+\partial_{\bar w}\cG\partial_w(\cA_+-\cA_-)}~,
    \nonumber\\
    C_{(2)}&=\frac{2i}{3}\left(
    \frac{\partial_{\bar w}\cG\partial_w\cA_++\partial_w \cG \partial_{\bar w}\bar\cA_-}{3\kappa^{2}T^2} - \bar{\mathcal{A}}_{-} - \mathcal{A}_{+}  \right) \vol_{S^2}~,
\end{align}
where $\vol_{S^2}$ and $ds^2_{S^2}$ are the volume form and line element for unit-radius $S^2$, 
and $ds^2_{AdS_6}$ is the line element of unit-radius $AdS_6$.
The composite quantities $\cG$, $\kappa^2$, $T$ and $R$ are defined by
\begin{align}\label{eq:kappa2-G}
	\kappa^2&=-|\partial_w \cA_+|^2+|\partial_w \cA_-|^2~,
	&
	\partial_w\cB&=\cA_+\partial_w \cA_- - \cA_-\partial_w\cA_+~,
	\nonumber\\
	\cG&=|\cA_+|^2-|\cA_-|^2+\cB+\bar{\cB}~,
	&
	T^2&=\left(\frac{1+R}{1-R}\right)^2=1+\frac{2|\partial_w\cG|^2}{3\kappa^2 \, \cG }~.
\end{align}
An explicit expression for $\cG$ in terms of single-valued dilogarithm functions was derived in Appendix~C of~\cite{Uhlemann:2020bek};
it will not be needed for our purposes.

\subsection{Free energy}\label{eq:F-extr}

The free energies can be obtained holographically from the on-shell action of Type IIB supergravity, which directly computes the (large $N$) SCFT partition function. This follows from the basic AdS/CFT dictionary and has been verified for the theories considered here in \cite{Fluder:2018chf,Fluder:2019szh,Uhlemann:2019ypp}.
The free energy can also be obtained by evaluating the entanglement entropy associated with a spherical region in flat space, which is computed holographically by evaluating the area of a minimal surface.
Both approaches were studied holographically in~\cite{Gutperle:2017tjo}, where general expressions were derived and it was shown that the two approaches agree for a set of examples.
The expression derived from the entanglement entropy computation reads~\cite{Gutperle:2017tjo}
\begin{align}\label{eq:F-0}
	F&=-\frac{64\pi}{9(2\pi\alpha^\prime)^4}\int_{\Sigma} d^2w \,\kappa^2\cG~.
\end{align}
For regular solutions $\kappa^2$ and $\cG$ are positive in the interior of $\Sigma$ and zero on the boundary, so that $F$ is negative.
In Appendix~\ref{sec:F-dev} we show that this expression agrees with the free energy derived from the on-shell action for general functions $\cA_\pm$ as given in equation (\ref{eqn:cA}).
We also evaluate the resulting expression for the free energy explicitly.
As a main result, we show in Appendix~\ref{sec:F-dev} that, assuming that the regularity conditions (\ref{eqn:constr}) are satisfied, 
the free energy evaluates to
\begin{align}\label{eq:F}
	F&=-\frac{64\pi^2}{9(2\pi\alpha^\prime)^4} \sum_{\ell,k,m,n=1}^L Z^{[\ell k]}Z^{[m n]}\cL_3\left(\frac{r_k-r_m}{r_k-r_\ell}\frac{r_\ell-r_n}{r_m-r_n}\right)~,
\end{align}
where $\cL_3$ is the single-valued trilogarithm function, defined as~\cite{Zagier2007}
\begin{align}\label{eq:cL3-def}
	\cL_3(z)&=\Re\left(\Li_3(z)-\ln |z| \cdot \Li_2(z)-\frac{1}{3}\ln^2\!|z|\cdot \ln(1-z)\right)~.
\end{align}
The function $\cL_3$ is real-analytic on $\CC$ except for $\lbrace 0,1,\infty\rbrace$, where it is still continuous
and takes the values $\cL_3(0)=\cL_3(\infty)=0$ and $\cL_3(1)=\zeta(3)$.
It satisfies a number of functional relations, among which we highlight $\cL_3(z)=\cL_3(1/z)$ and 
\begin{align}\label{eq:cL3-rel}
	\cL_3(z)+\cL_3(1-z)+\cL_3\left(\frac{z}{z-1}\right)&=\zeta(3)~.
\end{align}
The terms in equation (\ref{eq:F}) where not all of $(\ell,k,m,n)$ are distinct may be evaluated explicitly, leading to
\begin{align}\label{eq:cJ-split}
	F&=\frac{64\pi^2}{9(2\pi\alpha^\prime)^4}\Bigg[\zeta(3) \sum_{\ell,k=1}^L \Big(Z^{[\ell,k]}\Big)^2 - \sum_{\stackrel{k,\ell,m,n}{\text{distinct}}}Z^{[\ell k]}Z^{[m n]}\cL_3\left(\frac{r_k-r_m}{r_k-r_\ell}\frac{r_\ell-r_n}{r_m-r_n}\right)\Bigg]\,.
\end{align}
Since $Z^{[\ell,k]}$ is purely imaginary, the first term is non-positive.
These expressions are not entirely in terms of SCFT data yet, since they still involve the positions of the poles characterizing the supergravity duals.
For a given set of 5-brane charges characterizing a 5-brane junction, the pole positions still have to be determined from the regularity conditions in equation (\ref{eqn:constr}) and (\ref{eq:reg-zeros}).

\subsection{Extremality}

One may take an alternative perspective, and consider the expression for $F$ in (\ref{eq:F}) -- for given $Z_\pm^\ell$, \emph{i.e.}\ 5-brane charges -- as defining a function of the pole positions, $F_{\rm trial}(\vec{r})$ with  $\vec{r}=\lbrace r_\ell\rbrace_{\ell=1}^{L}$.
This ``trial free energy'' $F_{\rm trial}(\vec{r})$ by definition agrees with $F$ for pole configurations that satisfy the regularity conditions (\ref{eqn:constr}). 
Moreover, since $\cL_3$ is finite, $F_{\rm trial}(\vec{r})$ is well-defined for any choice of $\vec{r}$. 
The relation to the original expression in  (\ref{eq:F-0}), however, is lost:
since the regularity conditions (\ref{eqn:constr}) were used in deriving the expression in (\ref{eq:F}) from (\ref{eq:F-0}), the two functions do not have to agree when the regularity conditions in (\ref{eqn:constr}) are not satisfied, and the interpretation of the expression in (\ref{eq:F-0}) indeed becomes subtle.\footnote{The regularity conditions in (\ref{eqn:constr}) ensure that $\cG$ is constant along the boundary and can be made to vanish on $\partial\Sigma$ by a judicious choice of the integration constant that is unspecified by the definition in (\ref{eq:kappa2-G}). If (\ref{eqn:constr}) is not satisfied, $\cG$ can not be set to zero on the entire boundary, and the integration constant remains as free parameter. The divergences of $\kappa^2$ at the poles are less suppressed in the integrand as a result (though they remain integrable).\label{foot:div}}
We show in Appendix~\ref{app:F-extr} that the configurations of poles which satisfy the regularity conditions in (\ref{eqn:constr})
are distinguished points also in the parameter space of $F_{\rm trial}(\vec{r})$: 
if the conditions in (\ref{eqn:constr}) are satisfied, then
\begin{align}\label{eq:extr}
	\qquad \qquad
	\frac{\delta F_{{\rm trial}}(\vec{r})}{\delta r_\ell}=0~, \qquad \forall \ell =1,\ldots, L~.
\end{align}
That is, any configuration of poles satisfying the regularity conditions in (\ref{eqn:constr}) is a local extremum of the function $F_{\rm trial}(\vec{r})$.

One may wonder whether this can be turned around, \emph{i.e.} whether the regularity conditions (\ref{eqn:constr}) can be derived from requiring the trial free energy to be extremal.
We have not attempted a proof, but suspect it may be possible.
The extremality conditions (\ref{eq:extr}) are $L$ equations, but $F_{\rm trial}(\vec{r})$ only depends on the cross ratios formed out of the entries of $\vec{r}$, which are invariant under the $SL(2,\RR)$ automorphisms of the upper half plane. So there are three directions along which $F_{\rm trial}(\vec{r})$ is constant and only $L-3$ of the conditions in (\ref{eq:extr}) are non-trivial. 
This parallels the  discussion of the regularity conditions (\ref{eqn:constr}), which allow for three poles to be fixed arbitrarily and then determine the remaining ones up to discrete degeneracy.

Moving further in this direction, one may wonder whether the remaining constraint (\ref{eq:reg-zeros}), which requires all zeros of $\partial\cA_-$ to be in the lower half plane, may also imprint itself on the trial free energy.
Indeed, using the numerical methods described in Section~\ref{sec:numerics}, we tested for a sample of $10^3$ randomly generated solutions for each $L\in\lbrace 4,5,6,7,8\rbrace$ whether the critical point of $F_{\rm trial}(\vec{r})$ is a maximum, a minimum, or a saddle point. 
In all cases where (\ref{eqn:constr}) and (\ref{eq:reg-zeros}) are both satisfied it turned out to be a minimum of $F$/maximum of $-F$. More precisely, the matrix $\delta F_{\rm trial}(\vec{r})/(\delta r_p \delta r_q)$ has three eigenvalues which are zero, reflecting the freedom to make $SL(2,\RR)$ transformations on the upper half plane, while all the remaining $L-3$ eigenvalues were positive in all cases.\footnote{This statement becomes trivial for $L=3$, where also the constraint in (\ref{eq:reg-zeros}) becomes trivial.}
For all configurations satisfying the constraint in (\ref{eqn:constr}) but not the one in (\ref{eq:reg-zeros}), we found that the $L-3$ non-trivial eigenvalues had mixed signs. This suggests that the condition in (\ref{eq:reg-zeros}) is reflected in the trial free energy as well.
It would be interesting to understand whether one can uniquely identify the configuration of poles that satisfies the conditions in (\ref{eqn:constr}) and (\ref{eq:reg-zeros}) from properties of the trial free energy $F_{\rm trial}$ alone. We leave a more detailed discussion and geometric interpretation of the regularity conditions -- perhaps as a form of volume extremization --
for the future.

From a practical perspective, the extremality property (\ref{eq:extr}) is useful in that it simplifies the variation of the free energy due to an \emph{infinitesimal} change of the residues.
Changing the residues -- representing a deformation of the brane web, \emph{e.g.}\ corresponding to a relevant deformation of the SCFT -- in general induces a shift in the positions of the poles, due to the regularity conditions (\ref{eqn:constr}). 
The variation of the free energy is
\begin{align}
	\delta F&= \sum_{\ell=1}^L \left[\frac{\delta F}{\delta Z_+^\ell}\delta Z_+^\ell + \frac{\delta F}{\delta Z_-^\ell}\delta Z_-^\ell + \frac{\delta F}{\delta r_\ell}\delta r_\ell\right]\,.
\end{align}
However, thanks to the extremality property (\ref{eq:extr}), the change in the pole positions, $\delta r_{\ell}$, does not contribute to the variation of $F$.
Knowledge of how the poles adapt to an infinitesimal change in the residues is therefore not required to obtain the change in the free energy $\delta F$.
Using the symmetry under exchange of $(\ell k)$ and $(mn)$, one finds
\begin{align}\label{eq:delta-F}
	\delta F &=-
	\frac{128\pi^2}{9(2\pi\alpha^\prime)^4} \sum_{\ell, k,m,n=1}^L \, Z^{[\ell k]} \,
	\delta Z^{[m n]} \, \cL_3\left(\frac{r_k-r_m}{r_k-r_\ell}\frac{r_\ell-r_n}{r_m-r_n}\right)~.
\end{align}
This will be useful in studying the infinitesimal change in free energy between ``nearby'' fixed points.


\section{RG flows}\label{sec:flows}

From the large $N$ perspective, where the 5-brane charges are $\mathcal O(N)$, the flows discussed in Section~\ref{sec:5dscfts} may be divided into two categories: those corresponding to an $\mathcal O(N)$ change in the brane charges and those in which the brane charges change by a subleading amount.
From the perspective of the supergravity duals the latter correspond to an infinitesimal adjustment, while the former correspond to an $\mathcal O(1)$ change in the solution.

For instance, for the Higgs branch flows, removing a sub-junction involving $\mathcal O(1)$ 5-brane charges corresponds to an infinitesimal change in the supergravity dual, while removing a sub-junction involving $\mathcal O(N)$ charges leads to a substantially changed supergravity dual.
Similarly, for the mass deformations, if $K(p,q)$ and $K^\prime(p^\prime,q^\prime)$ are smaller than $\mathcal O(N)$ the deformation leads to an infinitesimal flow; if they are $\mathcal O(N)$ then the flow induces a finite change in the supergravity dual. In the following we denote by $\delta F$ and $\Delta F$ the changes in the free energy due to infinitesimal and finite RG flows in the aforementioned sense, respectively.

Some finite flows can be understood as successively removing small numbers of branes.
In that case, the change in free energy can be obtained by integrating small variations. 
An example is given by separating $K$ D5-branes from the $+_{N,M}$ junction in Figure~\ref{fig:plus-web}.
This can be understood as $K$ times removing a single D5-brane, and the total change in free energy can be obtained by integrating the variation $\delta F$ due to removal of a single D5-brane.
In general, this is not possible: 
When separating a sub-junction from a 5-brane junction, the sub-junction itself may contain light degrees of freedom that contribute to the IR free energy. An example is removing, at once, a full $+_{K,K}$ sub-junction with $K$ of $\mathcal O(N)$ from the $+_{N,M}$ theory.
Integrating the variations in the free energy due to consecutive removal of $+_{1,1}$ junctions would reproduce the free energy of the $+_{N-K,M-K}$ theory, but it would miss the contribution of the (decoupled) $+_{K,K}$ theory.
The two flows do not lead to the same IR fixed point.
We will consider both finite and infinitesimal flows in the following.

We discuss Higgs branch flows in Section~\ref{sec:higgs-flow} and $SU(2)_R$-preserving relevant deformations in Section~\ref{sec:mass-flow-sugra}.
For both we discuss simple flows analytically and sample more general flows numerically.
We generated a large number of supergravity solutions for $L\in\lbrace 3,\ldots,10\rbrace$ poles, the procedure for which is described in Section~\ref{sec:numerics} (some example solutions are provided in Appendix~\ref{app:sample}).
We evaluated the change in free energy due to finite and infinitesimal flows starting from these solutions and the results are discussed in Section~\ref{sec:results}.

\subsection{Higgs branch flows}\label{sec:higgs-flow}

We will discuss Higgs branch flows realized by separating a sub-junction of $K$ 5-brane stacks from a junction of $L$ 5-brane stacks describing the UV SCFT.
From the supergravity perspective, the condition for such a deformation to exist is that there is a set of $K$ poles whose residues sum to zero.
Without loss of generality, we can label the poles such that
\begin{align}
	\sum_{\ell=1}^K Z_\pm^\ell&=0~.
\end{align}
We will consider the IR SCFT resulting from the removal of a $K$-fold junction of 5-branes, leaving behind some fraction of the first $K$ 5-brane stacks of the original junction. The IR SCFT in general consists of two sectors: The first is the original $L$-fold junction with the charges of the first $K$ 5-branes reduced accordingly.
It is described by a supergravity solution with residues $\lbrace\tilde Z_+^\ell\rbrace_{\ell=1}^L$ given by 
\begin{align}\label{eq:HB-IR-res-1}
	\tilde Z_\pm^\ell&=\begin{cases}(1-\xi)Z_\pm^\ell ~, & \ell=1,\ldots, K~,  \\ Z_+^\ell~, & \ell=K+1,\ldots, L~,\end{cases}
\end{align}
where $0<\xi<1$.
For $K=2$ -- corresponding to separating an entire 5-brane -- this is the full IR theory; there are no light 5d degrees of freedom associated with the separated 5-brane.
For $K\geq 3$ the second sector corresponds to the separated $K$-junction, and is described by a supergravity solution with residues $\lbrace\hat Z_+^\ell\rbrace_{\ell=1}^K$ given by
\begin{align}\label{eq:HB-IR-res-2}
&&	\hat Z_+^\ell &= \xi Z_+^\ell~, & \ell&=1,\ldots,K~.
\end{align}
The free energy in the IR is the sum of the free energy for the two sectors.

For $K=2$, corresponding to removing an entire 5-brane from a junction which allows for that, \emph{i.e.}\ with $Z_\pm^1=-Z_\pm^2$, the minimal number of poles in the UV solution is $L=4$.
If two residues are opposite-equal, such that complete 5-branes can be removed, the remaining two residues are opposite-equal as well due to charge conservation.
The regularity conditions for four-pole solutions with two pairs of opposite-equal residues were discussed in~\cite{Gutperle:2017tjo}.
For solutions with $Z_+^1=-Z_+^2$ and $Z_+^3=-Z_+^4$ the poles can be chosen as 
\begin{align}
	r_1&=0~, & r_2&=\frac{2}{3}~, & r_3&=\frac{1}{2}~, & r_4 &=1~.
\end{align}
The free energy obtained from (\ref{eq:cJ-split}) is
\begin{align}\label{eq:HB-4-2}
	F_{\rm UV}&=\frac{28\zeta(3)}{3\pi^2 {\alpha^\prime}^4} \big(Z^{[1,3]}\big)^2~.
\end{align}
Since $Z^{[\ell,k]}$ is imaginary this is negative.
The free energy for the $+_{N,M}$ solution, as a particular example, has been matched to a localization computation in~\cite{Fluder:2018chf,Uhlemann:2019ypp}.
The IR SCFT consists of only one sector in this case, and is described by a 4-pole solution with pairwise opposite-equal residues as well.
The free energy with the residues (\ref{eq:HB-IR-res-1}) becomes
\begin{align}
	F_{\rm IR}&=\frac{28\zeta(3)}{3\pi^2 {\alpha^\prime}^4} \big(\tilde Z^{[1,3]}\big)^2 = (1-\xi)^2F_{\rm UV}~.
\end{align}
This clearly satisfies $F_{\rm IR}/F_{\rm UV}<1$, and (\ref{eqn:Fthm}), for $0<\xi<1$, as expected.

For $K=3$ -- \emph{i.e.} separating a triple junction -- the separated part of the 5-brane junction describes ungapped degrees of freedom. 
The minimal case is $L=K=3$, for which the free energy of the UV SCFT, obtained from (\ref{eq:cJ-split}), is
\begin{align}\label{eq:3-pole-F}
	F_{\rm UV}&=\frac{8\zeta(3)}{3\pi^2{\alpha^\prime}^4}(Z^{[1,2]})^2~.
\end{align}
Separating the triple junction into two triple junctions leaves two SCFTs that are decoupled at the leading order in large $N$.
One is described by a supergravity solution with residues $\tilde Z_+^\ell=(1-\xi) Z_+^\ell$, the other by a supergravity solution with residues $\hat Z_+^\ell=\xi Z_+^\ell$.
The IR free energy is given by
\begin{align}
	F_{\rm IR}&=(\xi^4+(1-\xi)^4)F_{\rm UV}~.
\end{align}
As expected, $F_{\rm IR}/F_{\rm UV}<1$ and (\ref{eqn:Fthm}) is satisfied.
For a solution with 4 poles a deformation removing a triple junction does not exist, since three residues summing to zero in the UV solution would force the remaining residue to zero by charge conservation. The minimal number of poles with which non-trivial deformations separating off a triple junction can be realized is 5.
This case and cases with $K>3$ will be treated numerically in Section~\ref{sec:numerics}.

An expression for the general variation of the free energy due to infinitesimal flows, separating a small sub-junction, may be obtained from (\ref{eq:delta-F}).
Using the antisymmetry of $Z^{[\ell k]}$, we find
\begin{align}\label{eq:delta-F-Higgs}
	\frac{\delta F}{\delta \xi} &=
	\frac{256\pi^2}{9(2\pi\alpha^\prime)^4} \sum_{\ell, k,m=1}^L  Z^{[\ell k]}
	\sum_{n=1}^K	Z^{[m n]}	
	\cL_3\left(\frac{r_k-r_m}{r_k-r_\ell}\frac{r_\ell-r_n}{r_m-r_n}\right)\,.
\end{align}
The contribution of the separated sub-junction is $\mathcal O(\xi^4)$ and sub-leading.
An $F$-theorem would imply that the variation in (\ref{eq:delta-F-Higgs}) is always positive if $(r_\ell,Z_+^\ell)_{\ell=1}^L$ satisfy the regularity conditions.

\subsection{Mass deformations}\label{sec:mass-flow-sugra}

We now turn to the $SU(2)_R$-preserving relevant deformations discussed from the brane perspective in Section~\ref{sec:mass-flow}.
In the supergravity setup, we pick integers $s,t\in \lbrace 1,\ldots L\rbrace$ and let $r_s$ and $r_t$ be the neighboring poles in the UV solution for which the residues are decreased in magnitude. For the flows discussed in Section~\ref{sec:mass-flow}, a new brane stack will be created in the original solution at a position $r_{L+1}$ on the real line which is between $r_s$ and $r_t$.
The resulting supergravity solution has $L+1$ poles with residues $\lbrace\tilde Z_+^\ell\rbrace_{\ell=1}^{L+1}$ given by
$\tilde Z_+^\ell=Z_+^\ell$ for $s,t\neq \ell\in\lbrace 1,\ldots, L\rbrace$ and 
\begin{align}\label{eq:mass-IR-res-1}
\tilde Z_+^s&=(1-\alpha)Z_+^s~, 
	&\tilde Z_+^t&=(1-\beta)Z_+^t~,
	&\tilde Z_+^{L+1}&=\alpha Z_+^s+\beta Z_+^t~,
\end{align}
with $0<\alpha,\beta<1$.
As discussed in Section~\ref{sec:mass-flow} for the $Y_N$ theory, the IR SCFT in general consists of two sectors. For the example in Figure~\ref{fig:YN-def} these are coupled by weakly gauging global symmetries; the gauge fields accomplishing that are subleading in number and do not contribute to the leading-order free energy.
The first sector of the IR SCFT is described by the setup in (\ref{eq:mass-IR-res-1}); the second sector is described by a 3-pole solution with residues $\lbrace \hat Z_+^\ell\rbrace_{\ell=1}^3$ given by
\begin{align}\label{eq:mass-IR-res-3}
	\hat Z_+^1&=\alpha Z_+^s~, & \hat Z_+^2&=\beta Z_+^t~, & Z_+^3&=-\alpha Z_+^s-\beta Z_+^t~.
\end{align}
The IR free energy is the sum of the free energies for the $(L+1)$-pole solution and the 3-pole solution, \emph{i.e.} $F_{\rm IR}=\tilde F_{\rm IR} ( \tilde Z^{\ell}_{+} )+\hat F_{\rm IR} ( \hat Z^{\ell}_{+} )$.

The smallest number of poles to consider in the initial configuration for a mass deformation is three.
An example is shown in Figure~\ref{fig:YN-def}.
Similar to the Higgs branch deformations discussed above, we will treat the minimal case analytically.
The free energy of the UV SCFT was given in (\ref{eq:3-pole-F}).
With the residues left arbitrary, we can fix $(s,t)=(1,2)$ without loss of generality; the other cases follow by permuting the residues.
For this flow, the first sector of the IR SCFT is a quadruple junction, described by a 4-pole supergravity solution with poles at $\tilde r_\ell$ with residues $\tilde Z_+^\ell$ given by
\begin{align}
	(\tilde Z_+^\ell)_{\ell=1}^4 &= \left( (1-\alpha)Z_+^1, (1-\beta)Z_+^2,Z_+^3,\alpha Z_+^1+\beta Z_+^2 \right)~,
	&
	(\tilde r_\ell)_{\ell=1}^4 &=\left( -1,0,1,u\right)~.
\end{align}
The poles corresponding to $s=1$ and $t=2$ are at $-1$ and $0$, respectively, so $u\in(-1,0)$.
Using the relations in (\ref{eq:cL3-rel}) and that the original three residues $Z_+^\ell$ sum to zero, we find the following expression for the contribution to the IR free energy
\begin{align}
	\tilde F_{\rm IR}(\tilde Z^{\ell}_{+})&=
	\frac{16 (1-\alpha) (1-\beta)(Z^{[1,2]})^2}{3 \pi ^2 {\alpha^\prime}^4}
	\left[\alpha  \cL_3\left(\frac{u-1}{2u}\right)+\beta 
	\cL_3\left(\frac{1-u}{1+u}\right)-(\alpha+\beta)\zeta(3)\right]
	\nonumber\\ &\hphantom{=}\ 
	+(1-\alpha \beta )^2 F_{\rm UV}~,
\end{align}
with $F_{\rm UV}$ as in (\ref{eq:3-pole-F}).
The second sector is the triple junction, described by a 3-pole supergravity solution with 
residues given in (\ref{eq:mass-IR-res-3}), whose free energy is given by 
\begin{align}
\hat F_{\rm IR}(\hat Z^{\ell}_{+})&=\alpha^2\beta^2 F_{\rm UV}~.
\end{align}
For $\Delta F = F_{\rm IR} - F_{\rm UV}$ we find
\begin{align}\label{eq:DeltaF-3-pole-mass}
	\frac{\Delta F}{F_{\rm UV}}&=
	2\alpha  \beta  (\alpha  \beta-1)+2 (1-\alpha) (1-\beta)
	\left[\alpha  \left(\frac{\cL_3\big(\frac{u-1}{2u}\big)}{\zeta (3)}-1\right)+\beta \left(
	\frac{\cL_3\big(\frac{1-u}{1+u}\big)}{\zeta (3)}-1\right)\right].
\end{align}
The first term is negative for arbitrary $0<\alpha,\beta<1$.
The second term consists of the square bracket with a coefficient which is positive.
In the square brackets, the two combinations of $\cL_3$ and $\zeta(3)$ multiplying $\alpha$ and $\beta$ are separately negative for arbitrary $u\in(0,1)$.
So $\Delta F/F$ is negative regardless of whether $u$ solves the regularity condition (\ref{eqn:constr}).

Nevertheless, it is instructive to discuss the condition determining $u$.
Since the original 3-pole solution is entirely specified in terms of $Z^{[1,2]}$, and the change in residues is also dictated by $Z_+^1$ and $Z_+^2$, the residues drop out of the regularity conditions (which can be taken in the form (\ref{eq:constr-4}), with $\cA_\pm^0$ eliminated), leaving
\begin{align}\label{eq:3-pole-mass-reg}
	\alpha\ln\left|\frac{2u}{1-u}\right|&=\beta\ln\left|\frac{1+u}{1-u}\right|~.
\end{align}
For any positive $\alpha$, $\beta$ there is a solution $u\in (-1,0)$.
Solutions outside of this interval do not satisfy the constraint (\ref{eq:reg-zeros}), which is required for a regular supergravity solution.
To survey all deformations one may alternatively fix $u\in(-1,0)$ and solve for $\beta$ in terms of $\alpha$ such that $\alpha,\beta\in(0,1)$.

Before moving on to numerically studying more general flows, we discuss infinitesimal deformations, \emph{i.e.} $|\alpha|,|\beta|\ll 1$.
The contribution to the free energy from the triple junction, \emph{i.e.} $\hat F(\hat Z^{\ell}_{+})$ in the above notation, is subleading  (in $\alpha, \beta$), and does not affect the infinitesimal variation of the free energy.
Due to the extremality property (\ref{eq:extr}), the variation of the pole positions drops out of the variation of the free energy.
However, one does need the location of the pole $\tilde r_{L+1}$ in the limit where the residue $\tilde Z_+^{L+1}$ vanishes.
Thus, one effectively starts with an $(L+1)$-pole solution with the residue $Z_+^{L+1}$ zero at the UV fixed point, and we set
\begin{align}
	\delta Z_+^s&=-\alpha Z_+^s~, & \delta Z_+^t &=-\beta Z_+^t~, & \delta Z_+^{L+1}=\alpha Z_+^s+\beta Z_+^t~.
\end{align}
For $\alpha,\beta\rightarrow 0$, the first $L$ regularity conditions (\ref{eqn:constr}) approach their unperturbed form; the $L$ original poles approach their positions in the UV solution.
The remaining condition resulting from the pole at $r_{L+1}$ reads
\begin{align}\label{eq:ru-reg}
	\cA_+^0 \delta Z_-^{L+1} - \cA_-^0 \delta Z_+^{L+1}+\sum_{\ell=1}^L \delta Z^{[\ell,L+1]}\ln|r_\ell-r_{L+1}|&=0~,
\end{align}
and determines where the new pole emerges.

The variation of the free energy becomes
\begin{align}\label{eq:deltaF-mass}
	\delta F =
	\frac{256\pi^2}{9(2\pi\alpha^\prime)^4} \sum_{\ell\neq k} Z^{[\ell k]}\sum_{n=1}^L
	\Bigg[&
	\alpha  Z^{[s,n]}\cL_3\left(\frac{r_k-r_s}{r_k-r_\ell}\frac{r_\ell-r_n}{r_s-r_n}\right)
	+\beta  Z^{[t,n]}\cL_3\left(\frac{r_k-r_t}{r_k-r_\ell}\frac{r_\ell-r_n}{r_t-r_n}\right)
	\nonumber \\
	&
	-\left(\alpha  Z^{[s,n]}+\beta  Z^{[t,n]}\right)	\cL_3\left(\frac{r_k-r_{L+1}}{r_n-r_{L+1}}\frac{r_\ell-r_n}{r_\ell-r_k}\right)
	\Bigg]\,.
\end{align}
Note that $r_{L+1}$ depends on $\alpha$, $\beta$ and on the properties of the original solution.
An $F$-theorem would imply that this variation decreases $-F$ for arbitrary $0<\alpha,\beta\ll 1$.

\subsection{Numerics}\label{sec:numerics}

In order to test (\ref{eqn:Fthm}) for more general flows, we generated a large random sample of supergravity solutions, by solving the constraints (\ref{eqn:constr}) and (\ref{eq:reg-zeros}) for randomly chosen configurations of residues, and evaluated the change in free energy between the UV and IR fixed points for the RG flows discussed in Section~\ref{sec:higgs-flow} and \ref{sec:mass-flow-sugra}. We have done so for finite flows, for which (generically) one solution has to be generated for the UV fixed point and two solutions for the IR fixed point, as well as infinitesimal flows. 
This was done separately for Higgs branch deformations, for which solutions were generated such that a sub-junction of a predetermined number of 5-branes can be removed, and for the $SU(2)_R$-preserving relevant deformations, where this constraint is not required.

To generate a solution, one has to determine, for a given set of residues, the positions of the poles.
Two of the regularity conditions in (\ref{eqn:constr}), say for $k=m$ and $k=n$, are solved for the constants $\cA_\pm^0$. The result is given in (\ref{eq:cA0-sol}).
The remaining regularity conditions are then given by (\ref{eq:constr-4}), which may be written (for fixed $k,m,n$) as 
\begin{align}\label{eq:constr-num}
	\sum_{\ell=1}^L Z^{[\ell,k]}Z^{[m,n]}\ln|r_\ell-r_k|+(\text{even permutations of $(k,m,n)$})&=0~.
\end{align}
With the conditions for $k=m$ and $k=n$ trivial and the sum of the left-hand side over $k$ vanishing, these are $L-3$ independent conditions.
Three poles can be fixed arbitrarily and the remaining ones can be determined by numerically solving the independent regularity conditions.

Numerical solutions can be found efficiently with an educated initial guess for the pole positions $\lbrace r_\ell\rbrace$.
We used the following: We map the upper half plane to the unit disc, and then superimpose the brane junction picture as in Figure~\ref{fig:junction}, where the 5-branes are at angles determined by their 5-brane charges.
The intersection of the external 5-branes with the boundary of the disc, mapped back to the upper half plane, then provides an initial guess for the position of the poles.
This leads to
\begin{align}\label{eq:r0-num}
	r_\ell^{(0)}&=f\left(\frac{Z_+^\ell}{|Z_+^\ell|}\right)~, \qquad \text{with} \qquad f(z)=i\frac{1+z}{1-z}~.
\end{align}
As three poles can be fixed arbitrarily, we take three of these initial guesses as actual pole positions, while the remaining pole positions are solved for numerically.

To avoid numerically degenerate solutions we impose that the poles are separated by $10^{-3}$ on the real line (which can be accomplished by an $SL(2,\RR)$ transformation). 
Once a solution to the regularity conditions (\ref{eqn:constr}) is found, we impose the remaining condition (\ref{eq:reg-zeros}),
by numerically computing the zeros of $\partial\cA_-$ and ensuring that they are all in the lower half plane.\footnote{If all zeros are all in the upper half plane one can take $\Sigma$ to be the lower half plane.}
In terms of the brane picture, the condition (\ref{eq:reg-zeros}) enforces that the ordering of the poles is compatible with the ordering of the 5-branes in the brane junction.
The initial guess in equation (\ref{eq:r0-num}) implements this ordering, but the numerical search may not preserve it.

\begin{figure}
	\begin{tabular}{c@{\hspace{20mm}}c}
		\includegraphics[width=0.35\linewidth]{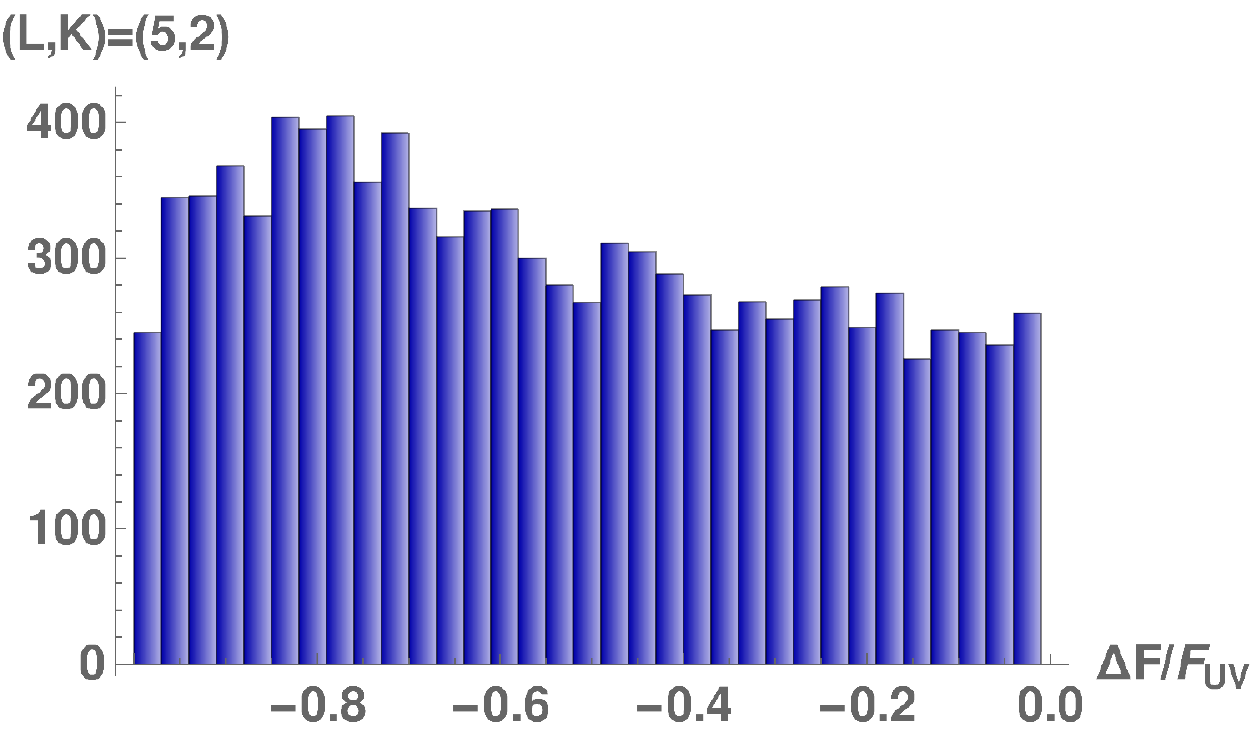}
		&
		\includegraphics[width=0.35\linewidth]{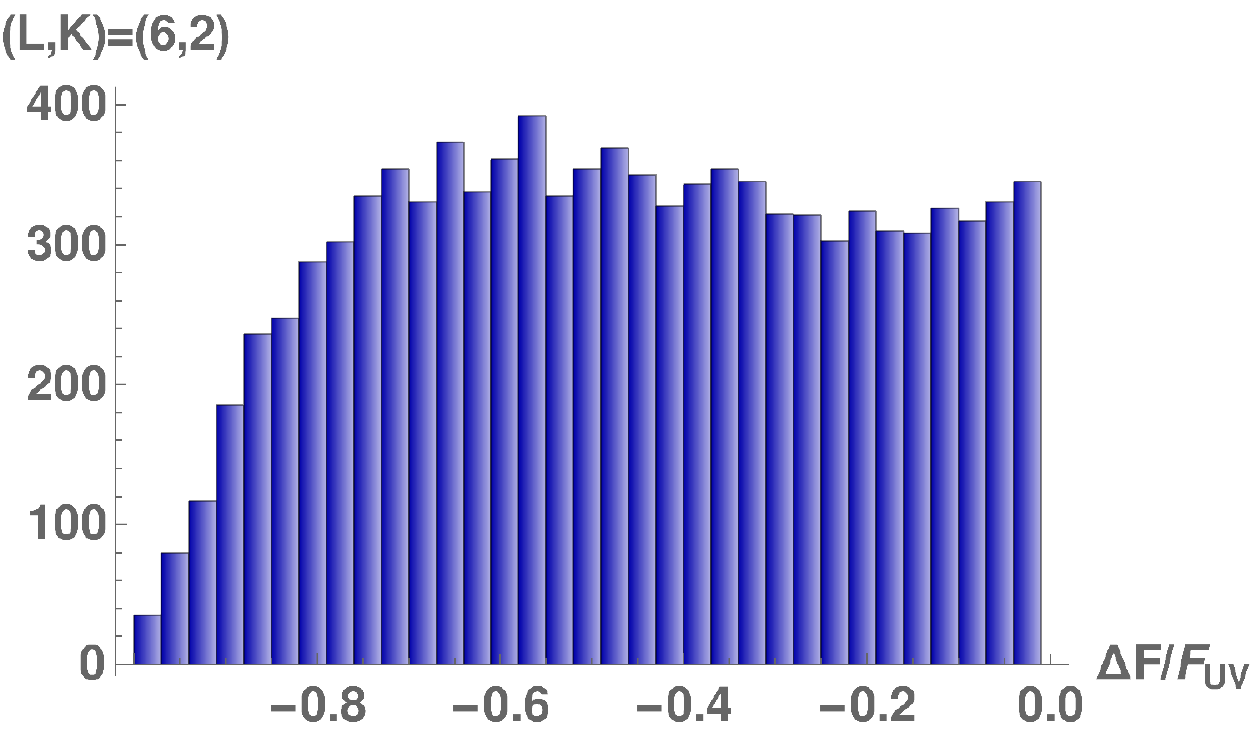}
		\\[5mm]
		\includegraphics[width=0.35\linewidth]{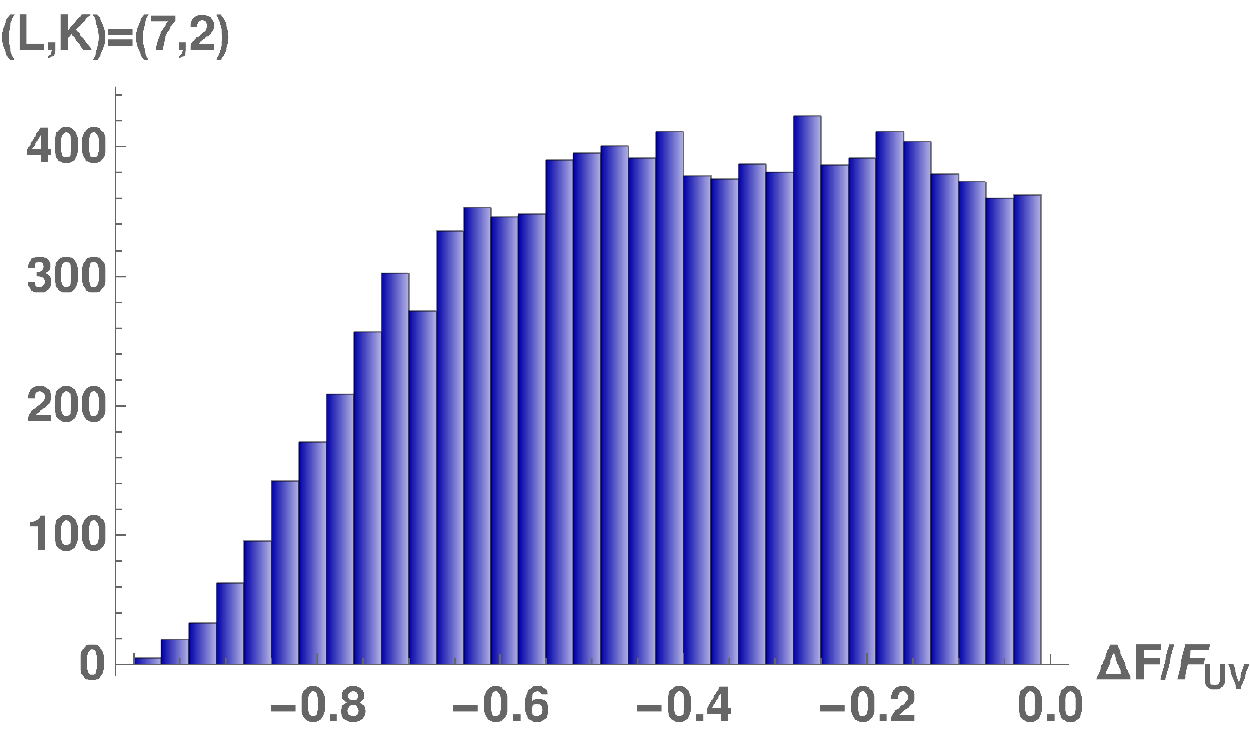}
		&
		\includegraphics[width=0.35\linewidth]{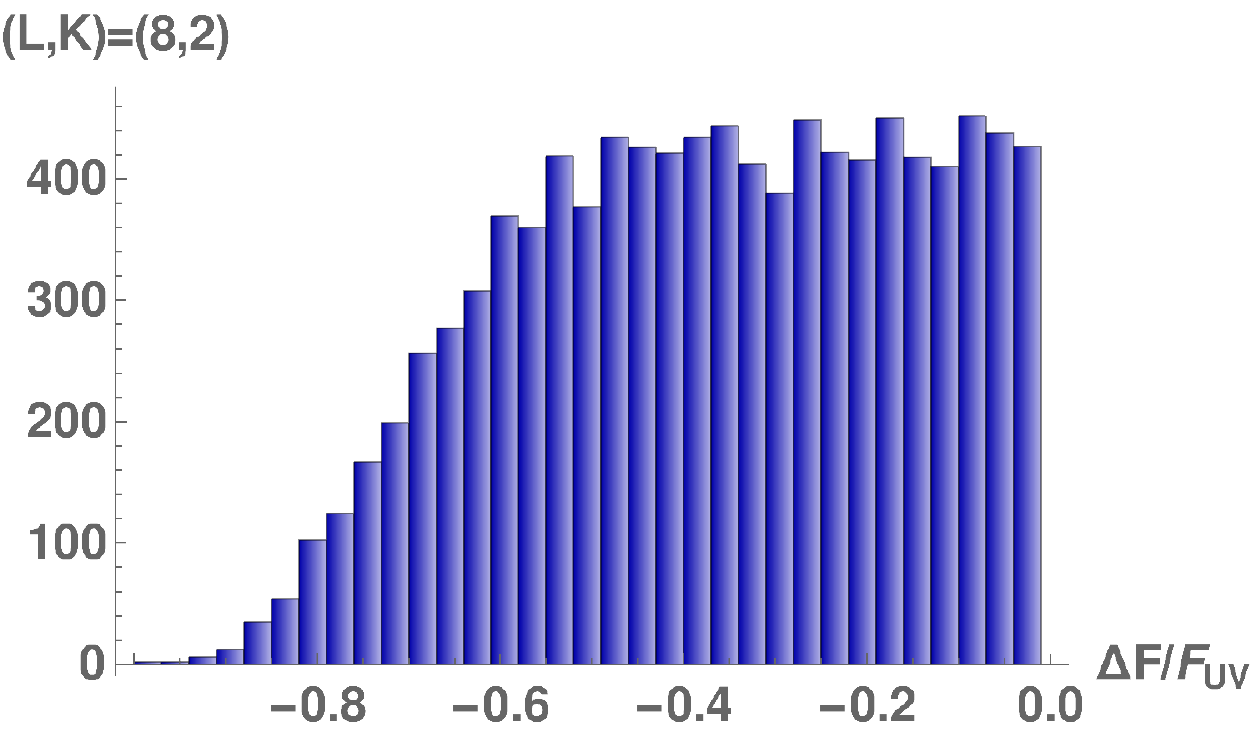}
	\end{tabular}
	
	\caption{Histograms for $\Delta F/F_{\rm UV}$ for Higgs branch flows removing an entire 5-brane, for $L=5,6,7,8$. For $L=9,10$ similar distributions were generated; they are qualitatively similar.\label{fig:HB2-hist}}
\end{figure}

\begin{figure}
	\begin{tabular}{c@{\hspace{20mm}}c}
		\includegraphics[width=0.35\linewidth]{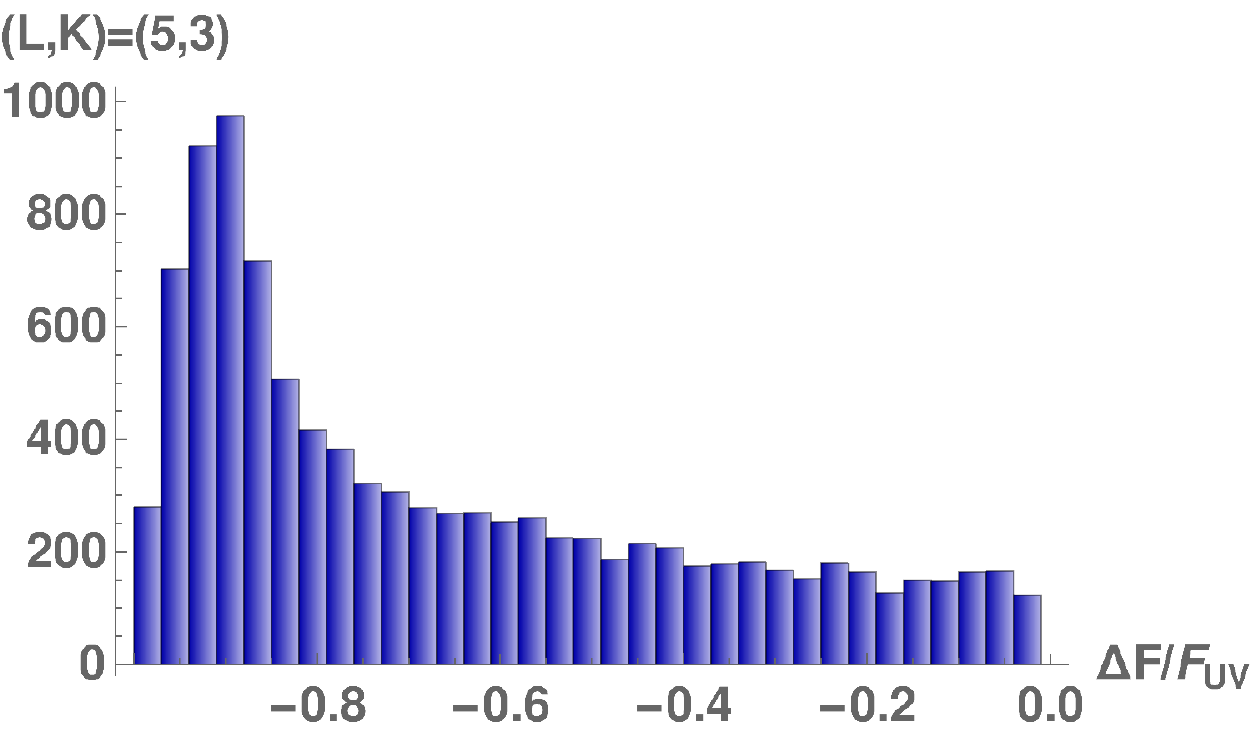}
		&
		\includegraphics[width=0.35\linewidth]{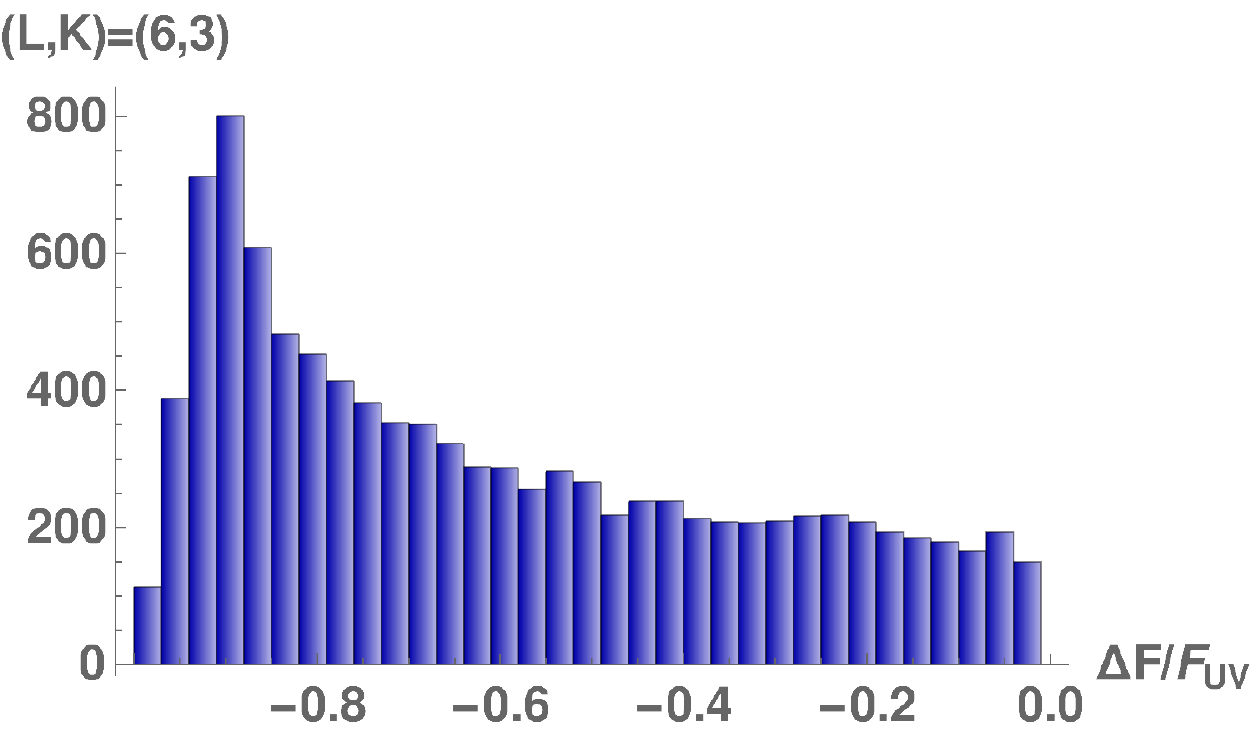}
		\\[5mm]
		\includegraphics[width=0.35\linewidth]{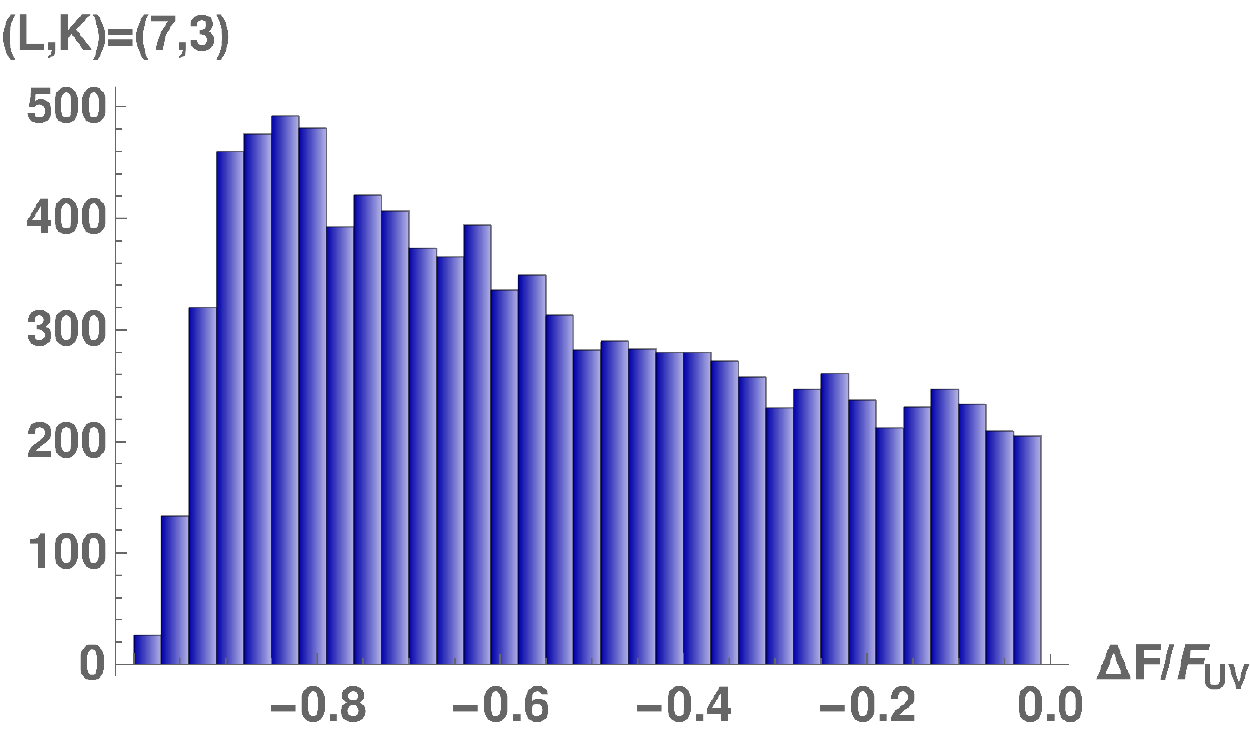}
		&
		\includegraphics[width=0.35\linewidth]{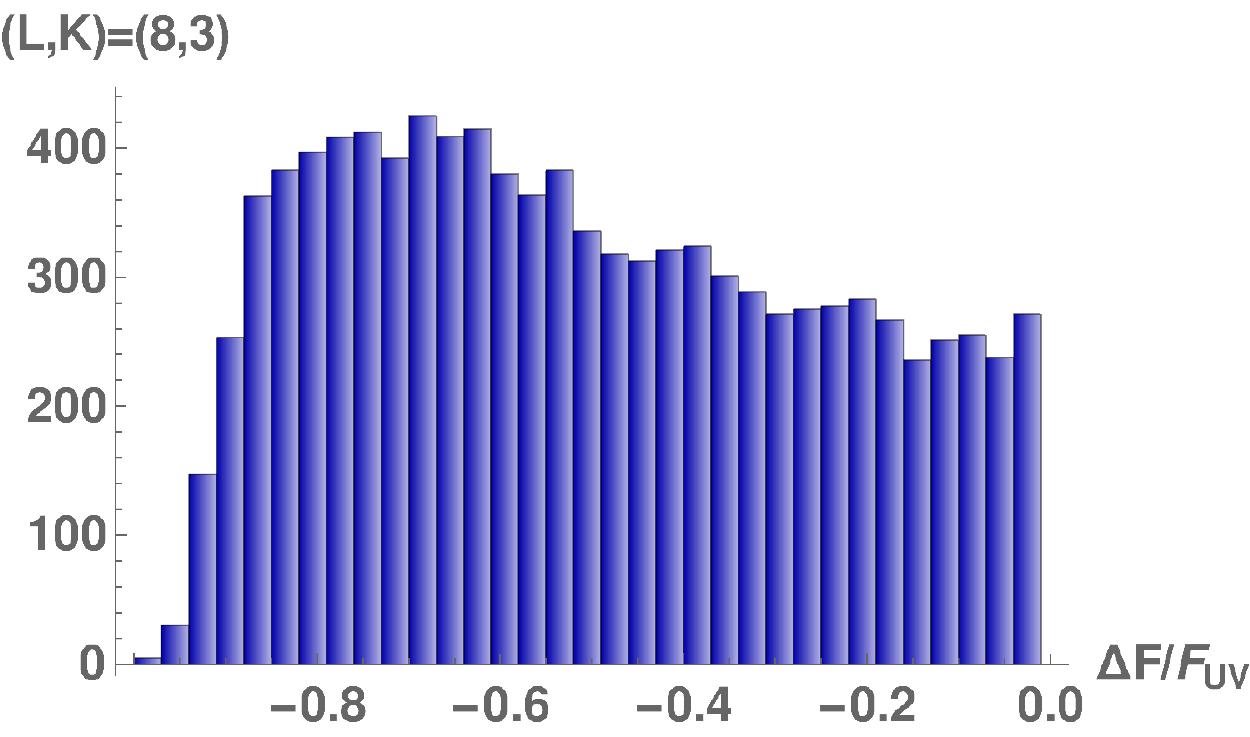}		
	\end{tabular}
	\caption{Histograms for $\Delta F/F_{\rm UV}$ for Higgs branch flows removing a triple junction, for $L=5,6,7,8$.
		The distributions for $L=9,10$ are again similar.\label{fig:HB3-hist}}
\end{figure}

\subsubsection{Results}\label{sec:results}

We start with the Higgs branch flows where a complete 5-brane is removed. 
Using the algorithm outlined in the previous section we generated a random sample of $10^4$ solutions for each $L\in \lbrace 5,6,7,8,9,10\rbrace$.
For each solution $L-1$ residues are drawn randomly in such a way that the first two are opposite-equal, and the remaining residue is fixed by charge conservation.
The pole positions are then determined numerically.
For each solution we evaluated the infinitesimal change in free energy following from (\ref{eq:delta-F-Higgs}).
The ratio $\delta F/F_{\rm UV}$ was negative for all solutions created.
As discussed in Section~\ref{sec:higgs-flow}, for these flows negativity of $\delta F/F_{\rm UV}$ is sufficient to conclude that finite flows are compatible with an $F$-theorem, and our data certainly suggests that $\delta F/F_{\rm UV}$ is generally negative.\footnote{%
If the regularity condition (\ref{eq:reg-zeros}) is dropped, however, to allow for irregular initial solutions, $\delta F/F$ can take either sign. Thus, a general proof of $\delta F/F<0$ has to incorporate (\ref{eqn:constr}) and (\ref{eq:reg-zeros}).}
We nevertheless generated the finite flows as well.
For each solution we chose a random $\xi\in(0,1)$ (see (\ref{eq:HB-IR-res-1})), and determined the $L$-pole supergravity solution corresponding to the IR fixed point. Since there is no contribution from the decoupled 5-branes, the IR free energy is given by the contribution from the solution with residues (\ref{eq:HB-IR-res-1}) alone.
The resulting data is visualized in Figure~\ref{fig:HB2-hist}.
The ratio $\Delta F/F_{\rm UV}$, with 
\begin{align}
\Delta F&=F_{\rm IR}-F_{\rm UV}~,
\end{align}
is negative throughout, in line with (\ref{eqn:Fthm}). 
As clearly exhibited in the histograms, 
$\Delta F/F_{\rm UV}$ is bounded from below by $-1$. As the number of poles is increased, the (negative) expectation value of $\Delta F/F_{\rm UV}$ shifts towards zero, which is to be expected: the impact of removing even an entire stack of $(p,q)$ 5-branes (corresponding to $\xi=1$) on the free energy naturally decreases as the number of brane stacks involved in the brane junction describing the UV SCFT increases.

\begin{figure}
	\begin{tabular}{c@{\hspace{20mm}}c}
		\includegraphics[width=0.35\linewidth]{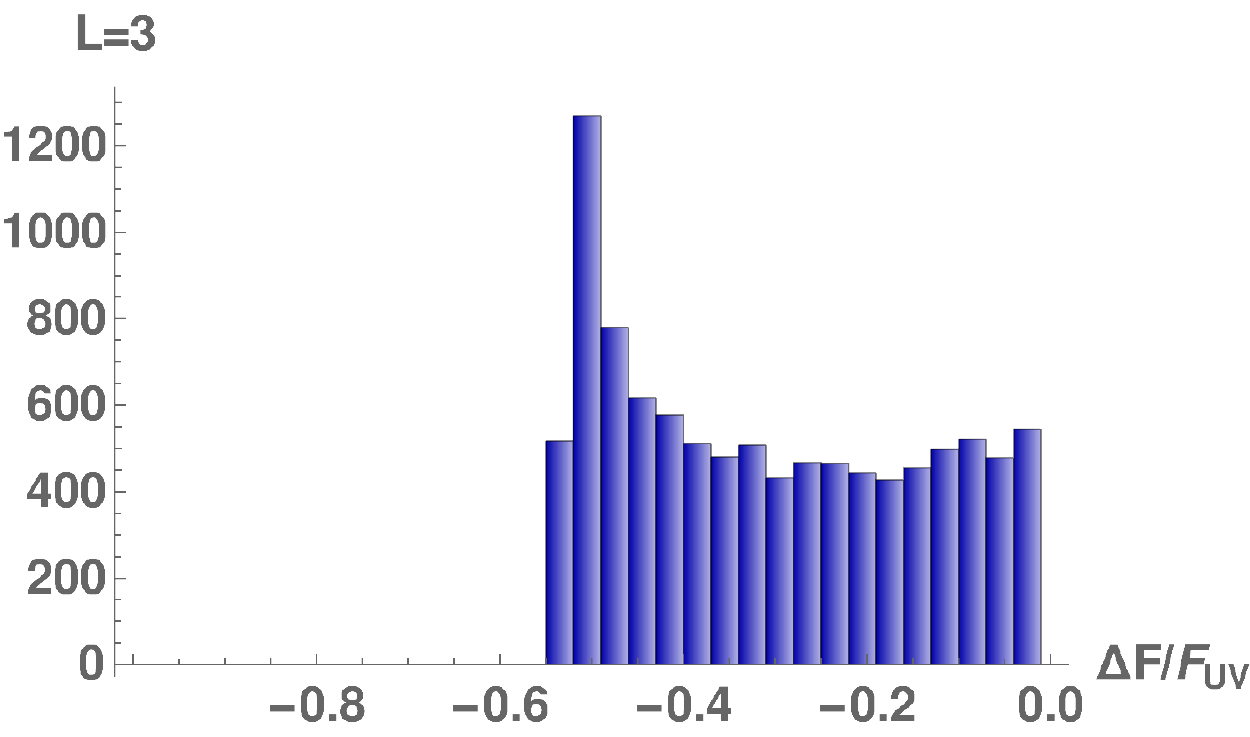}
		&
		\includegraphics[width=0.35\linewidth]{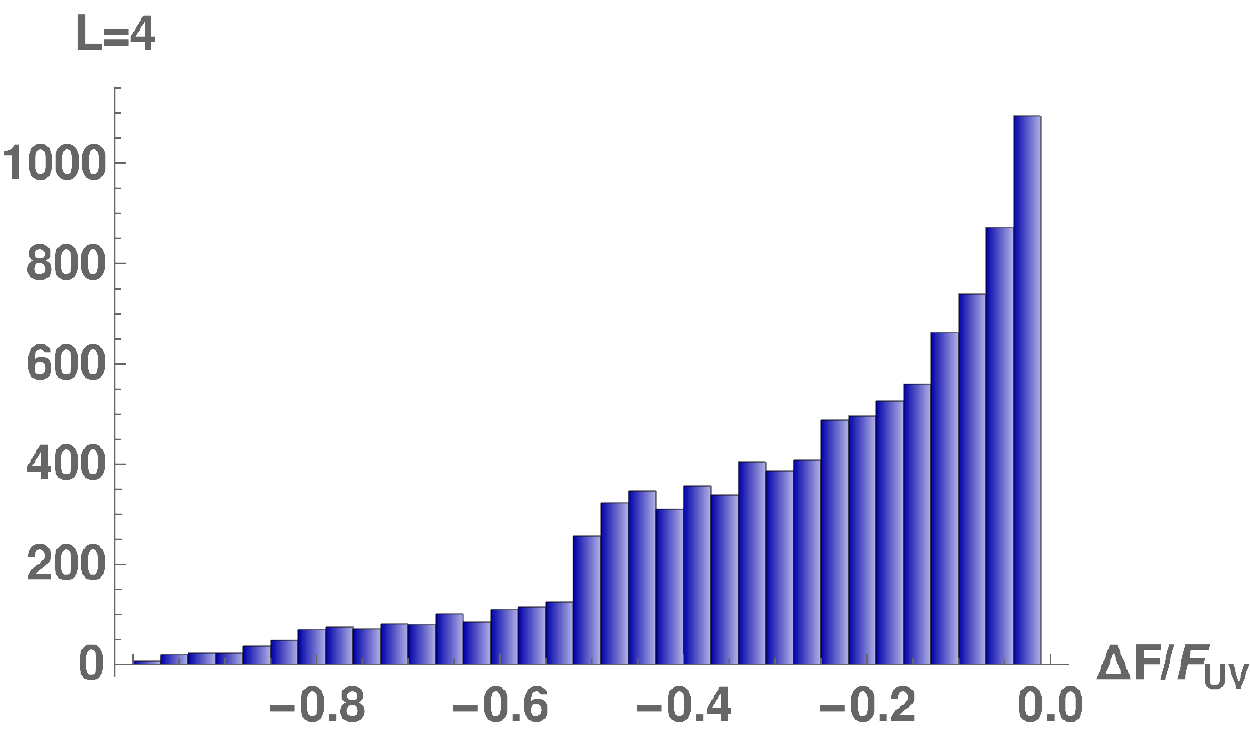}
		\\[5mm]
		\includegraphics[width=0.35\linewidth]{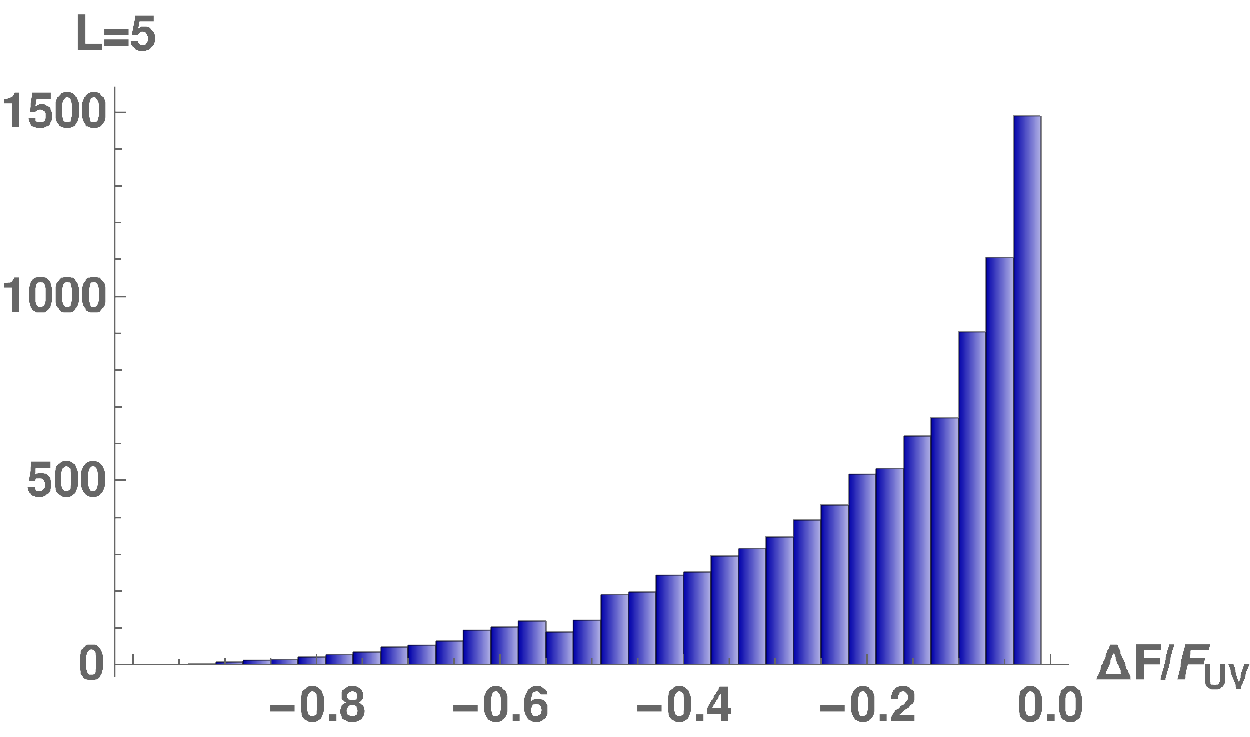}
		&
		\includegraphics[width=0.35\linewidth]{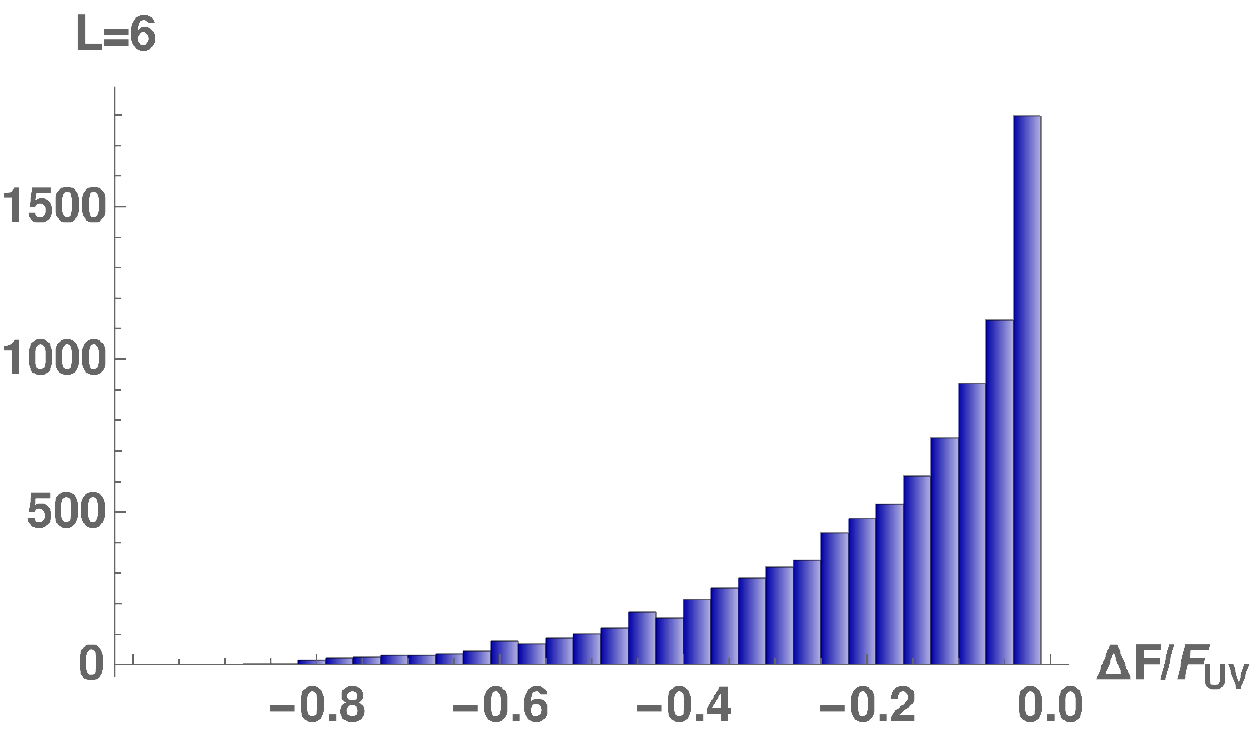}		
		\\[5mm]
		\includegraphics[width=0.35\linewidth]{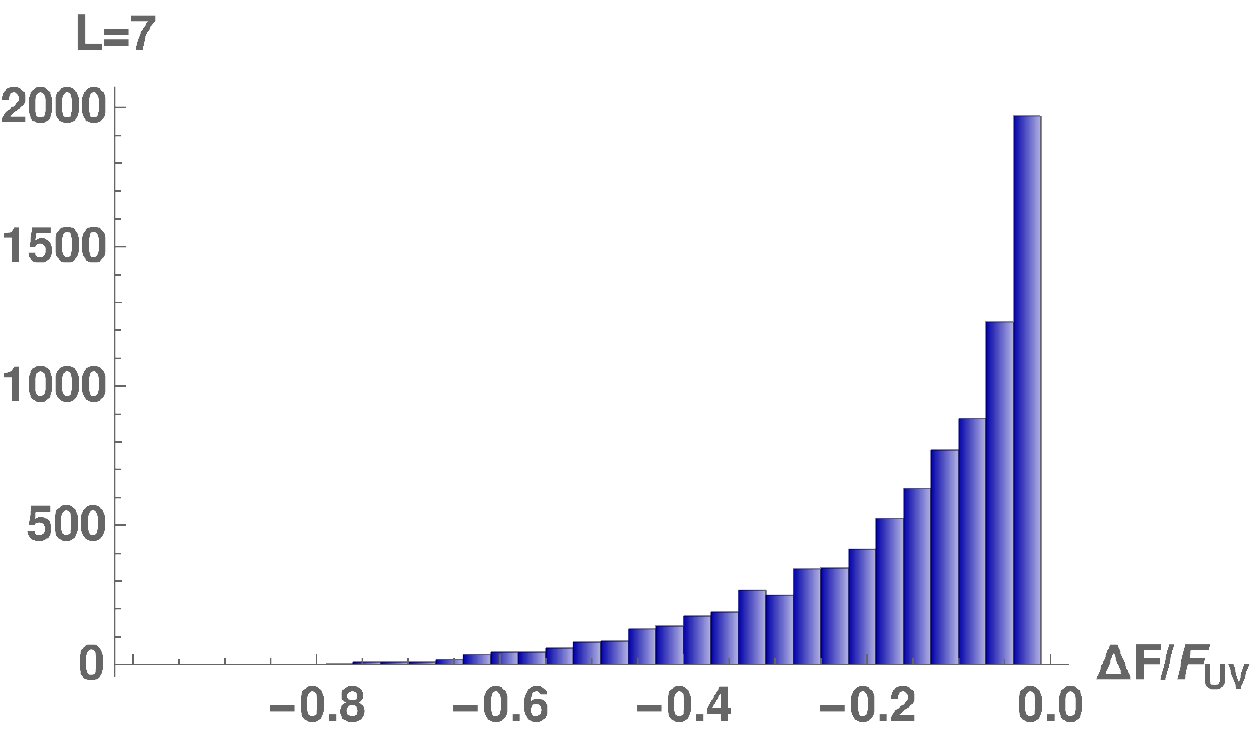}		
		&
		\includegraphics[width=0.35\linewidth]{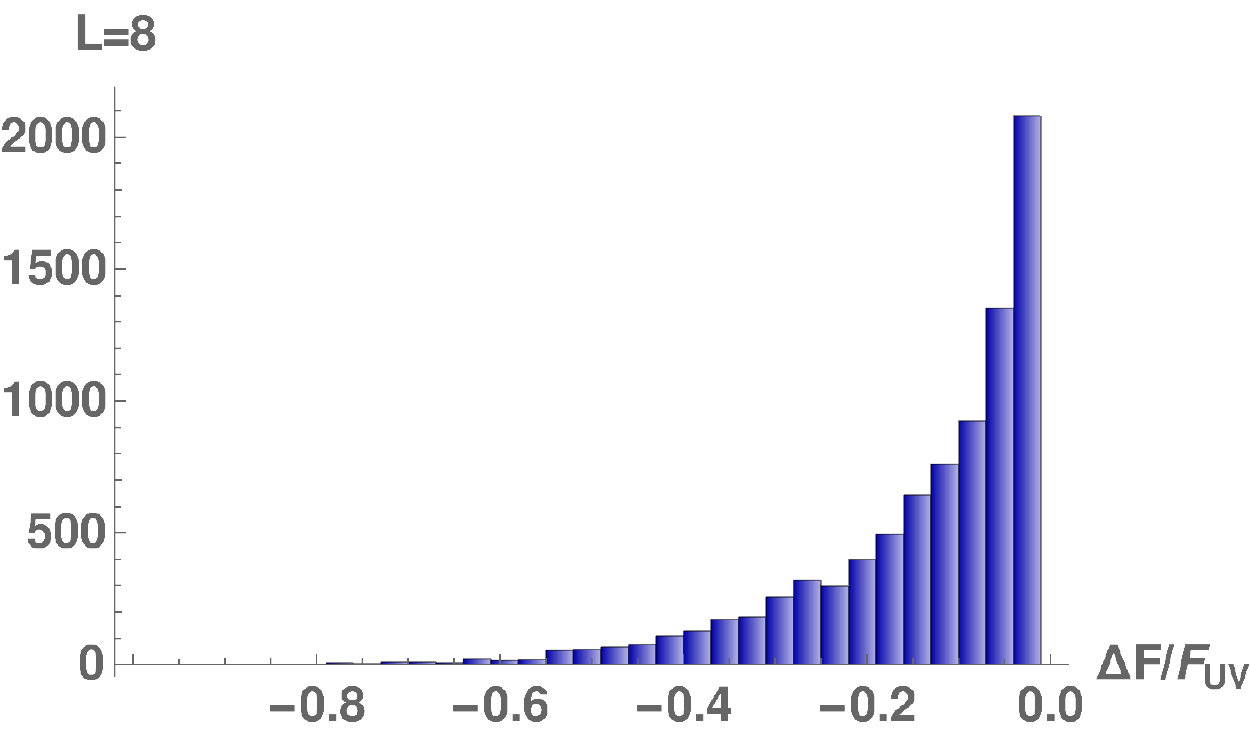}		
	\end{tabular}
	\caption{Histograms for $\Delta F/F_{\rm UV}$ for $SU(2)_R$-preserving mass deformations for $L=3,4,5,6$.
		For $L=3$ the lower bound is larger than $-1$, and agrees with the minimum of (\ref{eq:DeltaF-3-pole-mass}). 
		For larger $L$ the lower bound approaches $-1$. 
		The distributions for $L=7,8,9,10$ are qualitatively similar.
		\label{fig:mass-hist}}
\end{figure}

A similar survey was conducted for Higgs branch flows removing a triple junction. 
Using the algorithm outlined above, $10^4$ solutions were generated for each $L\in \lbrace 5,\ldots 10\rbrace$, such that three of the residues sum to zero and a Higgs branch flow can be realized by removing a triple junction.
We computed $\delta F/F$ using (\ref{eq:delta-F-Higgs}) with $K=3$ for each solution, and found it negative throughout.
We also analyzed finite flows.
For each solution we chose a random $\xi\in (0,1)$, specifying what fraction of the 5-branes constituting the 3-fold sub-junction should be separated off, and generated the $L$-pole solution with residues as in (\ref{eq:HB-IR-res-1}). 
For these flows the separated triple junction with residues as in (\ref{eq:HB-IR-res-2}) does contribute to the IR free energy and has to be taken into account.
We again found $\Delta F/F_{\rm UV}$ to be negative for all flows, in line with an $F$-theorem (\ref{eqn:Fthm}). The results are illustrated in Figure~\ref{fig:HB3-hist}.
As before, the expectation value of $\Delta F/F_{\rm UV}$ decreases in absolute value as the number of poles is increased, but it is negative for all solutions created.
We considered flows separating sub-junctions of $4$ and more brane stacks as well, with similar results.
The ratios $\delta F/F$ for infinitesimal flows and $\Delta F/F_{\rm UV}$ for finite flows were negative throughout.

We now turn to the $SU(2)_R$-preserving mass deformations discussed in Section~\ref{sec:mass-flow-sugra}. 
The solutions describing the UV fixed points can be generated with no restriction on the residues other than that the total charge is conserved,
and we again generated $10^4$ solutions for each $L\in\lbrace 3,\ldots, 10\rbrace$.
For the mass deformation one has a discrete choice of the two neighboring brane stacks ($s$ and $t$ in the notation of Section~\ref{sec:mass-flow-sugra}) and a choice of the parameters $\alpha$ and $\beta$ reflecting the fractions of branes in each of the two stacks that are separated from the junction.
For each solution we chose a deformation randomly and computed the infinitesimal change in the free energy from (\ref{eq:deltaF-mass}).
As discussed below (\ref{eq:3-pole-mass-reg}), it is actually more convenient to fix the position at which the new pole emerges in the correct interval and then determine $\beta$ in terms of $\alpha$. The variations $\delta F/F$ were all negative.
For finite flows, we again chose a deformation randomly.
In this case the contribution from the separated triple junction with residues (\ref{eq:mass-IR-res-3}) has to be taken into account, and the solution describing the leftover $(L+1)$-fold junction with residues (\ref{eq:mass-IR-res-1}) has to be generated.
The results for the change in free energy are shown in Figure~\ref{fig:mass-hist}.
For reference, the 3-pole case, which was treated analytically in Section~\ref{sec:mass-flow-sugra}, is included as well. 
Following the same argument as for the previous cases, $\Delta F/F_{\rm UV}$ is expected to decrease in absolute value as the number of poles is increased, which is borne out in the data.
The main point, of course, is that $\Delta F/F$ is negative throughout.


\section{Discussion}\label{sec:disc}

As reviewed in the introduction, $d=5$ stands out among the dimensions, in which superconformal field theories exist, in the scarcity of general results on $c$-functions that are available. Moreover, the sphere free energy $F$ as a candidate $c$-function is not without subtleties,
suggesting that a proper choice of assumptions may be crucial for establishing an $F$-theorem.

In this work, we have collected substantial evidence for the validity of an $F$-theorem for RG flows between 5d SCFTs engineered in Type IIB string theory by junctions of large numbers of $(p,q)$ 5-branes. These theories admit supergravity duals in Type IIB.
We have focused on the weak version of an $F$-theorem, expressed in equation~(\ref{eqn:Fthm}), and compared the sphere free energy between the end points of a large sample of RG flows.
The UV SCFTs were drawn from a random sample, and we considered flows triggered by relevant deformations that preserve the $SU(2)$ $R$-symmetry as well as Higgs branch flows. 
We studied $\mathcal O(10^5)$ RG flows numerically and found that in all cases the value of $-F$, computed from the holographic duals of the UV and IR SCFTs, was smaller at the IR fixed point than at the UV fixed point, in line with an $F$-theorem (\ref{eqn:Fthm}).
Moreover, for certain simple classes of RG flows we established analytically that $-F$ decreases and is compatible with an $F$-theorem.
These results certainly suggest that it should be possible to establish a general $F$-theorem for flows between 5d SCFTs with supergravity duals in Type IIB,
preferably in a form which also covers more general theories, \emph{e.g.} engineered by 5-brane junctions with 7-branes or orientifolds, for which supergravity solutions were discussed in~\cite{DHoker:2017zwj,Uhlemann:2019lge}.

We note that, for the theories considered here, the evidence for monotonic behavior of $-F$ directly extends to evidence for monotonicity of the conformal central charge $C_T$ which determines the stress-tensor three-point function (see \emph{e.g.} \cite{Chang:2017cdx,Chang:2017mxc}),
since $F$ is related to $C_T$ by a universal numerical factor at large $N$ \cite{Fluder:2018chf,Uhlemann:2019ypp}. 
While a $C_T$-theorem is not valid in 3d and 4d for theories with less than 8 supercharges (see \emph{e.g.} \cite{Nishioka:2013gza,Fei:2014yja}), supersymmetric theories in 5d have at least 8 supercharges, which may allow to establish a $C_T$-theorem.
It would be interesting to study the fate of the monotonicity properties beyond the strict large $N$ limit, to distinguish the two quantities.

From a more general perspective, we obtained new results for 5d SCFTs with holographic duals in Type IIB and found hints for interesting structures in their supergravity duals. We have shown that the free energy for a 5d SCFT engineered by a junction of 5-branes with charges $\lbrace (p_1,q_1),\ldots (p_L,q_L)\rbrace$, with all charges homogeneously large, is given by
\begin{align}\label{eq:F-discussion}
	F&=-\frac{64\pi^2}{9(2\pi\alpha^\prime)^4} \sum_{\ell,k,m,n=1}^L Z^{[\ell k]}Z^{[m n]}\cL_3\left(\frac{r_k-r_m}{r_k-r_\ell}\frac{r_\ell-r_n}{r_m-r_n}\right)~,
\end{align}
where $\cL_3$ is the single-valued trilogarithm function defined in (\ref{eq:cL3-def})
and  $Z^{[\ell k]}=Z_+^\ell Z_-^k-Z_+^k Z_-^\ell$ with $Z_\pm^\ell = \frac{3}{4}\alpha^\prime (\pm q_\ell + i p_\ell)$.
The parameters $r_1,\ldots, r_L$ are determined in terms of the 5-brane charges by the conditions in (\ref{eqn:constr}) and (\ref{eq:reg-zeros}).
These conditions arise from the requirement for regularity of the supergravity solutions and they are required for deriving (\ref{eq:F-discussion}).
However, one may change perspective and regard the expression in (\ref{eq:F-discussion}) as defining an ``off-shell" trial free energy for arbitrary $r_1,\ldots, r_L$. We have shown that the solutions to the conditions (\ref{eqn:constr}) extremize this trial free energy, and have given numerical evidence that the condition (\ref{eq:reg-zeros}) is satisfied if and only if the extremum is a maximum of $-F$.
This suggests that the entire set of supergravity regularity conditions may be recovered from an extremization principle for the trial free energy defined by (\ref{eq:F-discussion}), and hints at a geometric interpretation akin to a form of volume extremization.

For supergravity solutions that satisfy the regularity conditions, one can understand the volume computed by the free energy straightforwardly: 
From the entanglement entropy computations of $F$ in \cite{Gutperle:2017tjo}, one can identify a 4-manifold $M_4$ whose volume is the free energy. It takes the form of a warped product of $S^2$ over a disk/the upper half plane, with metric
\begin{align}\label{eq:metric-M4}
	ds^{2}_{M_4} = f_{6}^{2} \left( f_{2}^{2} d s^{2}_{S^{2}} + 4 \rho^{2} |d w|^2 \right)~,
\end{align}
where $f_6^2 f_2^2=2\cG/(3T)$ and $f_6^2\rho^2=\kappa^2 T$ with the definitions in (\ref{eq:kappa2-G}).
Our results indicate that the volume of this 4-manifold decreases between fixed points that can be connected by RG flows.
It would be interesting to understand how this interpretation would extend to holographic RG flow solutions, and whether there is a sense in which the metric on $M_4$ uniformizes along flows, \emph{e.g.} along the lines of \cite{Friedan:1980jm} where the Ricci flow equations were found in RG flows.

If the regularity conditions are not satisfied, the interpretation of $F$ in (\ref{eq:F-discussion}) as volume of $M_4$ becomes subtle (the metric in (\ref{eq:metric-M4}) acquires additional parameters and divergences, see footnote \ref{foot:div}).
However, the expression in (\ref{eq:F-discussion}) remains well-behaved and finite, suggesting that a geometric interpretation which differs from the volume of $M_4$ away from regular solutions may exist.
Intriguingly, $\cL_3$ features prominently in the volumes of hyperbolic 5-manifolds \cite{Goncharov:1996dce} (see also \cite{Filothodoros:2018pdj}).
It would be interesting to understand whether the expression in (\ref{eq:F-discussion}) may be related to triangulations of 5-manifolds.

We close with a more general outlook.
It would certainly be desirable to establish monotonicity results directly in field theory.
At least for planar theories it should be possible to make progress through explicit analyses.
For example, the general expression for the free energies of 5d SCFTs arising from balanced quiver gauge theories, derived in \cite{Uhlemann:2019ypp} from supersymmetric localization, allows to analyze fairly large classes of Higgs branch flows where multiple external 5-branes are terminated on the same 7-brane (see \cite{Benini:2009gi}) in field theory.
More generally, the identification of the saddle points for the matrix models resulting from supersymmetric localization with electrostatics potentials discussed in \cite{Uhlemann:2019ypp} may allow to cover more general flows and theories. 
One may hope to extend these methods to include corrections to the planar limit and perhaps gain useful insights into theories at finite $N$.
From a geometric perspective, it would be interesting to better understand the 5d free energy in terms of a volume maximization, akin to volume extremizations in other dimensions~\cite{Martelli:2005tp,Martelli:2006yb,Couzens:2018wnk,Gauntlett:2018dpc,Hosseini:2019use,Hosseini:2019ddy,Gauntlett:2019roi,Kim:2019umc}. The latter have had profound consequences and interpretations in the corresponding field theories, and it would be interesting to extend a similar logic to the present case.

\let\oldaddcontentsline\addcontentsline
\renewcommand{\addcontentsline}[3]{}
\begin{acknowledgments}
	The work of MF is supported by the SNSF fellowship P400P2-180740 and the Princeton Physics Department.
	CFU is supported, in part, by the US Department of Energy under Grant No.~DE-SC0007859 	and by the Leinweber Center for Theoretical Physics.
\end{acknowledgments}
\let\addcontentsline\oldaddcontentsline

\appendix
\renewcommand\theequation{\Alph{section}.\arabic{equation}}


\section{Free energy derivation}\label{sec:F-dev}

In the following, the expression for the free energy in (\ref{eq:F}) is derived.
We start with the free energy obtained from the entanglement entropy of a spherical region, and show that it matches the expression obtained from the on-shell action. We then proceed to evaluate it explicitly.

\let\oldaddcontentsline\addcontentsline
\renewcommand{\addcontentsline}[3]{}

\subsection{General expression}\label{sec:F-dev-1}

The free energy obtained from the entanglement entropy of a spherical region was discussed in~\cite{Gutperle:2017tjo}.
It is evaluated from an 8d minimal surface, which after using the form of the metric functions in~(\ref{eqn:ansatz}) and~(\ref{eq:kappa2-G}) leads to
\begin{align}\label{eq:F-A}
	F&=-\frac{64\pi}{9(2\pi\alpha^\prime)^4}\cJ~, &
	\cJ&=\int d^2w \,\left| \partial\cG\right|^2~.
\end{align}
From the definition of $\cG$ and the holomorphy of $\cB$, one finds that, without further assumptions,
\begin{align}\label{eq:dGsq-2}
	\left| \partial\cG\right|^2 \ = \ & \frac{1}{4}\partial\bar\partial\left(\cG^2-2(\cB+\bar\cB)\cG\right)+\bar\partial\left(\cG\partial\cB\right)+\partial\left(\cG\bar\partial\bar \cB\right)~.
\end{align}
The last two terms in (\ref{eq:dGsq-2}) produce boundary terms in the integral (\ref{eq:F-A}) which vanish due to  the regularity condition $\cG\vert_{\partial\Sigma}=0$. The $\cG^2$ term vanishes for the same reason.
Thus, for regular solutions,
\begin{align}\label{eq:cJ-3}
	\cJ&=- \frac{1}{2}\int_\Sigma d^2w\, \partial\bar\partial\left[(\cB+\bar\cB)\cG\right]~.
\end{align}
Using the definition of $\cG$ to replace $\cB+\bar\cB$ and again that $\partial\bar\partial\cG^2$ produces a boundary term which vanishes for regular solutions thanks to $\cG\vert_{\partial\Sigma}=0$, we can rewrite $\cJ$ as follows
\begin{align}
	\cJ&=\frac{1}{2}\int_\Sigma d^2w\, \partial\bar\partial\left[(|\cA_+|^2-|\cA_-|^2)\cG\right]~.
\end{align}
This expression holds for generic $\cA_\pm$, \emph{e.g.}\ also for solutions with 7-branes as in~\cite{DHoker:2017zwj,Uhlemann:2019lge}, as long as the regularity condition $\cG\vert_{\partial\Sigma}=0$ is satisfied.

\medskip

We now use the explicit form of $\cA_\pm$ for solutions describing 5-brane junctions in (\ref{eqn:cA}).
Take $\Sigma = \mathbb{H}^{+}$ the upper half plane, and conventions as in~\cite{Gutperle:2017tjo}, \emph{i.e.} $\partial_w=\frac{1}{2}\left(\partial_x-i\partial_y\right)$ and $d^2w=\frac{i}{2}dw\wedge d\bar w=dx\wedge dy$. Then, $\cJ$ can explicitly be written as follows
\begin{align}
	\cJ&=-\frac{i}{4}\int_{\partial\Sigma = \mathbb{R}} dx\, \partial\left[(|\cA_+|^2-|\cA_-|^2)\cG\right]~.
\end{align}
On the real axis, with the branch cut of $\ln$ along the negative real axis,
\begin{align}
	-|\cA_+|^2+|\cA_-|^2
	&=-2\pi i\sum_{\ell\neq k}Z^{[\ell,k]}\Theta\left(r_k-w\right)\ln\left|\frac{r_\ell-w}{r_\ell-r_k}\right|~.
\end{align}
Moreover, (see (C.6) of~\cite{Uhlemann:2020bek}),
\begin{align}
	\partial_w\cG
	&=\sum_{\stackrel{\ell,k=1}{\ell\neq k}}^L Z^{[\ell,k]} \frac{1}{w-r_k}\ln\left|\frac{w-r_\ell}{r_\ell-r_k}\right|^2~.
\end{align}
Consequently,
\begin{align}\label{eq:cJ-gen}
	\cJ&=\pi \sum_{\ell\neq k}\sum_{m\neq n}Z^{[\ell k]}Z^{[m n]}\int_{-\infty}^{r_k} dx\,
	\ln\left|\frac{x-r_\ell}{r_\ell-r_k}\right|
	\frac{1}{x-r_n}\ln\left|\frac{x-r_m}{r_m-r_n}\right|~.
\end{align}
This result agrees with the on-shell action of Type IIB supergravity (\emph{cf.}\ equation~(47) in~\cite{Gutperle:2017tjo}).

The lower bound in the integration can be shifted to an arbitrary position so long as it is independent of $r_\ell$ and $r_k$:
To see this, split $\ln\big|\frac{x-r_\ell}{r_\ell-r_k}\big|=\ln|x-r_\ell|-\ln|r_\ell-r_k|$.
This splits the extra integral that needs to be added to shift the integration bound into two pieces.
The first one, with $\ln|x-r_\ell|$, is independent of $r_k$, and hence vanishes when summed over $k$.
The second one depends on $r_\ell$ and $r_k$ only through the overall factor $\ln|r_\ell-r_k|$, which is symmetric under exchange of $r_\ell$ and $r_k$.
So it vanishes upon summing over $k$ and $\ell$,
showing that the lower bound can be shifted.

\subsection{Integration}

To explicitly evaluate $\cJ$ it will be convenient to introduce
\begin{align}
	y&=\frac{r_m-x}{r_m-r_n}~, & t_\ell&=\frac{r_m-r_\ell}{r_m-r_n}~, & t_k&=\frac{r_m-r_k}{r_m-r_n}~.
\end{align}
Then, upon shifting the lower bound of integration in (\ref{eq:cJ-gen}) to $r_n$, $\cJ$ can be written as
\begin{align}\label{eq:cJ-gen-change}
	\cJ&=\pi \sum_{\ell\neq k}\sum_{m\neq n}Z^{[\ell k]}Z^{[m n]}\int_{0}^{t_k} dy\,
	\ln\left|\frac{y-t_\ell}{t_\ell-t_k}\right| \frac{1}{y-1}\,\ln\left|y\right|~.
\end{align}

Due to $Z^{[\ell,k]}$ being antisymmetric in $\ell$ and $k$, the expression for $\cJ$ in (\ref{eq:cJ-gen}) is (anti)symmetric under various permutations of $(r_\ell,r_k,r_m,r_n)$, which will be useful below.
In the new variables, $(y, t_{\ell}, t_{k})$, these permutations are realized as follows:
Exchanging $r_n$ and $r_m$ corresponds to $\lbrace y,t_\ell,t_k\rbrace\leftrightarrow \lbrace 1-y,1-t_\ell,1-t_k\rbrace$;
exchanging $r_k$ and $r_\ell$ amounts to exchanging $t_\ell$ and $t_k$.
The remaining symmetry, \emph{i.e.} exchanging $(r_m,r_n)$ with $(r_\ell,r_k)$, under which the sum in $\cJ$ is symmetric, acts
on $t_\ell$, $t_k$ as follows
\begin{align}\label{eq:mn-lk-swap}
	t_\ell&\leftrightarrow t_\ell^\prime=\frac{t_\ell}{t_\ell-t_k}~,
	&
	t_k&\leftrightarrow t_k^\prime= \frac{t_\ell-1}{t_\ell-t_k}~.
\end{align}

One can use $\ln|y|/(y-1)=-\Re\partial_y\Li_2(1-y)$ and integration by parts 
to express $\cJ$ in terms of multiple  polylogarithms of weight three. The space of multiple polylogarithms of weigth three is spanned by regular polylogarithms of weight up to three~\cite{Duhr:2011zq}, and we shall now rewrite $\cJ$ in terms of these simpler functions.

The integral can be expressed in terms of the single-valued trilogarithm function~\cite{Zagier2007}
\begin{align}
	\cL_3(z)&=\Re\left(\Li_3(z)-\ln |z| \cdot \Li_2(z)-\frac{1}{3}\ln^2\!|z|\ln(1-z)\right)~.
\end{align}
This function is real-analytic except at $ z \in \lbrace 0,1,\infty\rbrace$, where it is continuous and evaluates to $\cL_3(0)=\cL_3(\infty)=0$ and $\cL_3(1)=\zeta(3)$.
Furthermore, the derivative along the real line is given by
\begin{align}\label{eq:dL3}
	\partial_x\cL_3(x)&=\frac{1}{3}\ln|x|\left(\frac{\ln|x|}{1-x}+\frac{\ln|1-x|}{x}\right)~.
\end{align}
Finally, we will also use that, on the real line,
\begin{align}
	\partial_x \Re\Li_2(x)&=-\frac{\ln|1-x|}{x}~.
\end{align}

Now, let us define
\begin{align}\label{eq:cD}
	\cD(y) \ = \ & \cL_3\left(\frac{t_\ell-y}{(t_\ell-1)y}\right)-\cL_3\left(\frac{t_\ell-y}{t_\ell-1}\right)-\cL_3(y)-\cL_3\left(1-\frac{t_\ell}{y}\right)
	+\frac{1}{3}\ln|1-y|\ln|y|\ln|y-t_\ell|
	\nonumber\\ &
	+\ln|t_k-t_\ell| \Re\Li_2(1-y)+\ln|t_\ell-1|\Re\left(\Li_2(y)-\Li_2\left(\frac{y}{t_\ell}\right)\right)
	\nonumber\\ &
	+\frac{1}{3} \ln|t_\ell-1| \left(\ln|t_\ell| \ln|t_\ell-y|+2\ln|y|\ln\left|\frac{t_\ell (1-y)}{t_\ell-y}\right|\right)~.
\end{align}
The real part of $\Li_2$ is single-valued and so is $\cD$.
Moreover, $\cD$ is continuous at $y=0$ and $y=t_\ell$.
An explicit computation shows that
\begin{align}
	\partial_y \cD(y)&=\ln\left|\frac{y-t_\ell}{t_\ell-t_k}\right| \frac{1}{y-1}\,\ln\left|y\right|~.
\end{align}
Therefore, we can express $\cJ$ as
\begin{align}
	\cJ&=\pi \sum_{\ell\neq k}\sum_{m\neq n}Z^{[\ell k]}Z^{[m n]}
	\big[\cD(t_k)-\cD(0)\big]~.
\end{align}
Since $\sum_k {Z_\pm^k}=0$, terms which are independent of $t_k$ vanish when summed over $k$.
The terms in the first and third line of (\ref{eq:cD}) are continuous at $y=0$, with no dependence on $t_k$. Hence, they drop out when evaluated at $y=0$.
The first term in the second line is symmetric under exchange of $t_\ell$ and $t_k$, and thus drops out. Lastly, the second term is independent of $t_k$.
Thus, together we may drop the $\cD(0)$-piece and end up with
\begin{align}
	\cJ&=\pi \sum_{\ell\neq k}\sum_{m\neq n}Z^{[\ell k]}Z^{[m n]}	\cD(t_k)~.
\end{align}

We can now evaluate the contribution from $y=t_k$ more explicitly. Noting that $(t_\ell-t_k)/(t_\ell-1)$ is independent of $r_m$, $t_k$ is independent of $r_\ell$ and $1-t_\ell/t_k$ is independent of $r_n$, only the first of the four $\cL_3$ terms survives summation.
Hence, we remain with the following expression for $\cJ$
\begin{align}
	\cJ \ = \ \pi \sum_{\ell\neq k}\sum_{m\neq n}Z^{[\ell k]}Z^{[m n]}
	\Bigg[&
	\cL_3\left(\frac{t_\ell-t_k}{(t_\ell-1)t_k}\right)
	+\frac{1}{3}\ln|1-t_k|\ln|t_k|\ln|t_k-t_\ell|
	\nonumber\\ &
	+\ln|t_k-t_\ell| \Re\Li_2(1-t_k)+\ln|t_\ell-1|\Re\left(\Li_2(t_k)-\Li_2\left(\frac{t_k}{t_\ell}\right)\right)
	\nonumber\\ &
	+\frac{1}{3} \ln|t_\ell-1| \left(\ln|t_\ell| \ln|t_\ell-t_k|+2\ln|t_k|\ln\left|\frac{t_\ell (1-t_k)}{t_\ell-t_k}\right|\right)\Bigg].
\end{align}
The second term in the first line and the first term in the last line together are symmetric under exchange of $t_\ell$ and $t_k$;
they therefore cancel upon taking the sum.
Exchanging $r_n$ and $r_m$ corresponds to $\lbrace t_\ell,t_k\rbrace \rightarrow \lbrace 1-t_\ell,1-t_k\rbrace$.
This can be used to merge the $\Li_2(t_k)$ and $\Li_2(1-t_k)$ terms.
Finally, we swap $\ell$ and $k$ in the last term, at the expense of a sign, to arrive at
\begin{align}\label{eq:cJ-cJ1-cJ2}
	\cJ \ = \ \pi \sum_{\ell\neq k}\sum_{m\neq n}Z^{[\ell k]}Z^{[m n]}\cL_3\left(\frac{t_\ell-t_k}{(t_\ell-1)t_k}\right)
	+\pi\left(\tilde \cJ_1+\frac{2}{3}\tilde\cJ_2\right)\,,
\end{align}
where
\begin{align}\label{eq:cJ1-cJ2}
	\tilde \cJ_1&=\sum_{\ell\neq k}\sum_{m\neq n}Z^{[\ell k]}Z^{[m n]}
	\Re\left(
	\ln\left|\frac{t_\ell-1}{t_k-t_\ell}\right| \,\Li_2(t_k)-\ln|t_\ell-1|\,\Li_2\left(\frac{t_k}{t_\ell}\right)
	\right)\,,
	\nonumber\\
	\tilde \cJ_2&=\sum_{\ell\neq k}\sum_{m\neq n}Z^{[\ell k]}Z^{[m n]} \ln\left|\frac{t_k-t_\ell}{(t_\ell-1)t_k}\right|  \ln|t_\ell|\ln|1-t_k|~.
\end{align}
The first term in $\cJ$ in equation (\ref{eq:cJ-cJ1-cJ2}), expressed in terms of $r_\ell$ and using that $\cL_3(1/z)=\cL(z)$, is identical to (\ref{eq:F}).
The restriction to $\ell\neq k$ and $m\neq n$ can be dropped, since $\cL_3$ vanishes in those cases.
To show that the free energy is indeed given by (\ref{eq:F}) it thus remains to show that $\tilde J_1=\tilde J_2=0$.

\subsection{Evaluating \texorpdfstring{$\tilde \cJ_1$ and $\tilde \cJ_2$}{cJ1 and cJ2}}

In general $\tilde \cJ_{1}$ and  $\tilde \cJ_{2}$ are non-zero. However, we shall now show that they in fact vanish when the poles satisfy the regularity conditions.

We start by evaluating $\tilde \cJ_2$; expressing $\tilde \cJ_2$ in terms of the pole positions leads to 
\begin{align}
	\tilde \cJ_2&=\sum_{\ell\neq k}\sum_{m\neq n}Z^{[\ell k]}Z^{[m n]} \ln\left|\frac{r_k-r_\ell}{r_\ell-r_n}\frac{r_m-r_n}{r_k-r_m}\right|  \ln\left|\frac{r_\ell-r_m}{r_m-r_n}\right|\ln\left|\frac{r_k-r_n}{r_m-r_n}\right|~.
\end{align}
One can restrict the sum to all of $\ell,k,m,n$ being distinct, since the product of logarithms vanishes otherwise.
Using the symmetries under exchange of $\ell \leftrightarrow k$, $m \leftrightarrow n$, as well as $(\ell,k) \leftrightarrow (m,n)$ allows to rewrite this as
\begin{align}
	\tilde \cJ_2&=2\sum_{\stackrel{\ell,k,m,n}{\text{distinct}}}
	Z^{[\ell k]}Z^{[m n]} L_{kn}\left(L_{\ell m}L_{k \ell}+L_{\ell n}L_{k m}-L_{m n}L_{k \ell}\right)~, \quad \text{where} \quad L_{ij}=\ln|r_i-r_j|
\end{align}
The last two terms are antisymmetric under the exchange of $k \leftrightarrow n$.
Thus, we may write
\begin{align}\label{eq:cJ2-f}
	\tilde \cJ_2&=2\sum_{\stackrel{\ell,k,m,n}{\text{distinct}}}
	Z^{[\ell k]}Z^{[m n]} L_{k n} L_{\ell m}L_{k \ell}+
	2\sum_{\stackrel{\ell,k,m,n}{\text{distinct}}}
	\left(Z^{[\ell k]}Z^{[m n]} - Z^{[\ell n]}Z^{[m k]}\right) L_{kn} L_{\ell n}L_{k m}~.
\end{align}
From the definition of $Z^{[\ell k]}$ we immediately obtain the identity 
\begin{align}\label{eq:ZZid}
Z^{[\ell k]}Z^{[m n]} - Z^{[\ell n]}Z^{[m k]}=Z^{[\ell m]}Z^{[k n]}~.
\end{align}
Using it in the second term of (\ref{eq:cJ2-f}) shows that, upon renaming indices $(\ell,n)\rightarrow (n,\ell)$ and then swapping $\ell$ and $k$ at the expense of a sign, 
it is opposite-equal to the first term. Thus, we have shown that
\begin{align}
\tilde \cJ_2=0~.
\end{align}

Now, let us turn to $\tilde \cJ_1$; when expressing it as in equation (\ref{eq:cJ1-cJ2}), in terms of the pole positions, it is convenient to add terms that are independent of $r_\ell$ and $r_\ell$, and which thus sum to zero. Thus, we can rewrite $\tilde \cJ_{1}$ it as follows
\begin{align}
	\tilde \cJ_1&=\sum_{\ell\neq k}\sum_{m\neq n}Z^{[\ell k]}Z^{[m n]}
	\ln\left|\frac{r_k-r_\ell}{r_n-r_\ell}\frac{r_m-r_n}{r_m-r_k}\right|\Re\left(
	\Li_2\left(\frac{r_m-r_k}{r_m-r_\ell}\right) - \Li_2\left(\frac{r_m-r_k}{r_m-r_n}\right)
	\right)\,.
\end{align}
If any of $(\ell,k,m,n)$ are equal their contribution vanishes, so the sum can be restricted to $(\ell,k,m,n)$ being distinct.
One can then pull the logarithm into the parenthesis and relabel indices $(\ell\leftrightarrow n)$ in the second term, to arrive at
\begin{align}\label{eq:cJ1-cI}
	\tilde \cJ_1&=\sum_{\stackrel{\ell,k,m}{\text{distinct}}} \cI_{\ell k m}\Re\Li_2\left(\frac{r_m-r_k}{r_m-r_\ell}\right)~,
\nonumber\\
	\cI_{\ell k m}&=\sum_{n\notin \lbrace \ell,k,m \rbrace}
	\left(Z^{[\ell k]}Z^{[m n]} \ln\left|\frac{r_k-r_\ell}{r_n-r_\ell}\frac{r_m-r_n}{r_m-r_k}\right|
	-Z^{[nk]}Z^{[m\ell]}\ln\left|\frac{r_k-r_n}{r_n-r_\ell}\frac{r_m-r_\ell}{r_m-r_k}\right|\right)~.
\end{align}

In the remainder of this section, we will show that $\cI_{\ell k m}$ vanishes if the regularity conditions (\ref{eqn:constr}) are satisfied.
To this end, we first bring the regularity conditions (\ref{eqn:constr}) into a useful form.
Pick two arbitrary but fixed conditions, $k=m$ and $k=n$, $m\neq n$, and solve for $\cA_\pm^0$. This yields
\begin{align}\label{eq:cA0-sol}
	Z^{[m,n]}\cA_\pm^0&=Z_\pm^n\sum_{\ell\neq m}Z^{[\ell,m]}\ln|r_\ell-r_m|-Z_\pm^m\sum_{\ell\neq n}Z^{[\ell,n]}\ln|r_\ell-r_n|~.
\end{align}
Then, using this in the regularity conditions yields
\begin{align}\label{eq:constr-4}
	\sum_{\ell\neq m} Z^{[\ell,m]}Z^{[n,k]}\ln|r_\ell-r_m|-\sum_{\ell\neq n} Z^{[\ell,n]}Z^{[m,k]}\ln|r_\ell-r_n|+\sum_{\ell\neq k}Z^{[\ell,k]}Z^{[m,n]}\ln|r_\ell-r_k|&=0~.
\end{align}
For $k=m,n$ these are trivial. So we have $L-3$ conditions, fixing $L-3$ of the poles. 

\medskip

Now, given (\ref{eq:constr-4}), let us evaluate $\cI_{\ell k m}$.
The terms in (\ref{eq:cJ1-cI}) which are independent of $r_n$ can be converted to (minus) sums over $n\in \lbrace \ell,k,m\rbrace$.
This leads to
\begin{align}
	\cI_{\ell k m} \ = \ &\sum_{n\notin \lbrace \ell,k,m \rbrace}
	\left(Z^{[\ell k]}Z^{[m n]} \ln\left|\frac{r_m-r_n}{r_n-r_\ell}\right|
	-Z^{[nk]}Z^{[m\ell]}\ln\left|\frac{r_k-r_n}{r_n-r_\ell}\right|\right)
	\nonumber\\
	&-\sum_{n\in\lbrace\ell,k,m\rbrace}
	\left(Z^{[\ell k]}Z^{[m n]} \ln\left|\frac{r_k-r_\ell}{r_m-r_k}\right|
	-Z^{[nk]}Z^{[m\ell]}\ln\left|\frac{r_m-r_\ell}{r_m-r_k}\right|\right)~.
\end{align}
Regrouping terms in the first line and using the identity (\ref{eq:ZZid}), we end up with
\begin{align}
	\cI_{\ell k m} \ = \ &\sum_{n\notin \lbrace \ell,k,m \rbrace}
	\Bigg(Z^{[\ell k]}Z^{[m n]} \ln\left|r_m-r_n\right|
	-Z^{[nk]}Z^{[m\ell]}\ln\left|r_k-r_n\right|
	-Z^{[km]}Z^{[n\ell]}\ln|r_n-r_\ell|
	\Bigg)
	\nonumber\\
	&-Z^{[\ell k]}Z^{[m\ell]}\ln\left|\frac{r_k-r_\ell}{r_m-r_\ell}\right|
	-Z^{[\ell k]} Z^{[mk]}\ln\left|\frac{r_k-r_\ell}{r_m-r_k}\right|
	+Z^{[mk]}Z^{[m\ell]}\ln\left|\frac{r_m-r_\ell}{r_m-r_k}\right|~.
\end{align}
The terms in the second line are exactly the terms excluded in the sums in the first line, so that
\begin{align}\label{eq:cIlkm}
	\cI_{\ell k m} \ = \ &\sum_{n}
	\left(Z^{[\ell k]}Z^{[m n]} \ln\left|r_m-r_n\right|
	-Z^{[nk]}Z^{[m\ell]}\ln\left|r_k-r_n\right|
	-Z^{[km]}Z^{[n\ell]}\ln|r_n-r_\ell|
	\right)\,,
\end{align}
where it is again understood that $Z^{[\ell k]}\ln |r_k-r_\ell|$ is zero if $\ell=k$.
With the regularity conditions as in equation~(\ref{eq:constr-4}), we conclude that
the right hand side vanishes.
In view of (\ref{eq:cJ1-cI}) this means that
\begin{align}
\tilde \cJ_1=0~.
\end{align}

\let\addcontentsline\oldaddcontentsline


\section{Extremality of \texorpdfstring{$F_{\rm trial}(\vec{r})$}{F}}\label{app:F-extr}

We start with the expression for $F$ where the sum is restricted to all summation indices being distinct -- as in equation (\ref{eq:cJ-split}) --
and translate it to $\cJ$ by stripping off overall factors -- as in equation (\ref{eq:F-A}).
Then, we consider the variation of $\cJ$ upon varying the position of a particular pole, $r_p$, \emph{i.e.}
\begin{align}
	\frac{\delta \cJ}{\delta r_p}&=
	\pi \!\sum_{\stackrel{k,\ell,m,n}{\text{distinct}}}\! Z^{[\ell k]} Z^{[m n]}\left[
	\delta_{m,p} \frac{\delta}{\delta r_m}+\delta_{n,p} \frac{\delta}{\delta r_n}
	+\delta_{\ell,p} \frac{\delta}{\delta r_\ell}+\delta_{k,p} \frac{\delta}{\delta r_k}
	\right]\cL_3\left(\frac{r_k-r_m}{r_k-r_\ell}\frac{r_\ell-r_n}{r_m-r_n}\right)\,.
\end{align}
The argument of $\cL_3$ is symmetric under exchange of $(\ell,k) \leftrightarrow (m,n)$, so the last two terms in the square brackets can be combined with the first two  terms. This leads to
\begin{align}
	\frac{\delta \cJ}{\delta r_p}&=
	2\pi \!\sum_{\stackrel{k,\ell,m,n}{\text{distinct}}}\! Z^{[\ell k]} Z^{[m n]} 
	\cL_3^\prime\left(\frac{r_k-r_m}{r_k-r_\ell}\frac{r_\ell-r_n}{r_m-r_n}\right)
	\left(\delta_{n,p}\frac{(r_k-r_m)(r_\ell-r_m)}{{(r_k-r_\ell)(r_m-r_n)^2}}-(m\leftrightarrow n)\right)\,.
\end{align}
The derivative of $\cL_3$ is given explicitly in (\ref{eq:dL3}), which leads to
\begin{align}
	\frac{\delta \cJ}{\delta r_p}&=
	\frac{2\pi}{3}\!\!
	\sum_{\stackrel{k,\ell,m,n}{\text{distinct}}}\!\! Z^{[\ell k]} Z^{[m n]} 
	\ln\left|x\right|
	\left[ \frac{\ln\left|x\right|}{1-x} +(m\leftrightarrow n)\right]
	\left[\delta_{n,p}\frac{(r_k-r_m)(r_\ell-r_m)}{{(r_k-r_\ell)(r_m-r_n)^2}}-(m\leftrightarrow n)\right],
\end{align}
where $x=(r_k-r_m)(r_\ell-r_n)/((r_k-r_\ell)(r_m-r_n))$.
The sum symmetrizes the $\ln|x|$ term in $(m,n)$, so one can replace $\ln|x|\rightarrow \frac{1}{2}(\ln|x|+(m\leftrightarrow n))$.
Furthermore, the ``$-(m\leftrightarrow n)$''-pieces in the last factor can then be absorbed into a factor 2.
This leads to
\begin{align}
	\frac{\delta \cJ}{\delta r_p}&=
	\frac{2\pi}{3}
	\sum_{\stackrel{k,\ell,m,n}{\text{distinct}}}\! Z^{[\ell k]} Z^{[m n]} \,
	\left[\ln\left|x\right|+(m\leftrightarrow n)\right]
	\left[\ln\left|x\right|\frac{r_m-r_k}{(r_k-r_n)(r_m-r_n)}-(k\leftrightarrow \ell)\right]
	\delta_{n,p}\,.
\end{align}
The first factor is symmetric in $k\leftrightarrow \ell$, so $-(k\leftrightarrow \ell)$ in the second factor only produces a factor 2, and we can rewrite
\begin{align}\label{eq:deltaJ-4}
	\frac{\delta \cJ}{\delta r_p}&=
	\frac{4\pi}{3}
	\!\sum_{\stackrel{(k,\ell,m)\neq p}{\text{distinct}}}\! Z^{[\ell k]} Z^{[m p]} 
	\ln\left|y(1-y)\right|
	\left(\frac{\ln\left|y\right|}{r_p-r_m}-\frac{\ln\left|y\right|}{r_p-r_k}\right),
	&
	y&=\frac{r_k-r_m}{r_k-r_\ell}\frac{r_\ell-r_p}{r_m-r_p}\,.
\end{align}
For the $1/(r_p-r_m)$ term, one can use that $\ln|y|\ln|y-1|$ is symmetric in $k\leftrightarrow \ell$, to reduce the coefficient to $\ln^2|y|$.
We find
\begin{align}
	&\sum_{\stackrel{(k,\ell,m)\neq p}{\text{distinct}}}\! Z^{[\ell k]} Z^{[m p]} \,
	\ln\left|y(1-y)\right|
	\frac{\ln\left|y\right|}{r_p-r_m}
	\nonumber\\
	&=
	\sum_{\stackrel{(k,\ell,m)\neq p}{\text{distinct}}}\! \frac{Z^{[\ell k]} Z^{[m p]}}{r_p-r_m}
	\left( 2L_{km}(L_{\ell p}-L_{k\ell})-2L_{k\ell}L_{\ell p}
	+2L_{mp}(L_{kp}-L_{km})-L_{kp}^2+L_{km}^2 \right)\,,
\end{align}
where as before $L_{ij}=\ln|r_i-r_j|$.
Now, we swap $k$ and $m$ in the second term and use the result in (\ref{eq:deltaJ-4}) to obtain
\begin{align}
	\frac{\delta \cJ}{\delta r_p}&=
	\frac{2\pi}{3}
	\sum_{m\neq p}\! \frac{1}{r_p-r_m}\hat\cJ_{pm}~,
\end{align}
where
\begin{align}
	\hat\cJ_{pm}=  
	\sum_{\stackrel{(\ell,k)\neq m,p}{\text{distinct}}}\Big[&
	Z^{[\ell k]} Z^{[m p]} \left( 2L_{km}(L_{\ell p}-L_{k\ell})-2L_{k\ell}L_{\ell p} +2L_{mp}(L_{kp}-L_{km})-L_{kp}^2+L_{km}^2 \right)
	\nonumber\\[-3mm]
	& +Z^{[\ell m]} Z^{[k p]} \, 
	(L_{m\ell}-L_{km}-L_{\ell p}+L_{kp}) (L_{mp}+L_{k\ell}+L_{m\ell}+L_{km}+L_{kp}+L_{\ell p})
	\nonumber\\
	& -3Z^{[\ell m]} Z^{[k p]} \, 
	(L_{m\ell}-L_{km}-L_{\ell p}+L_{kp}) (L_{m\ell}+L_{kp})
	\Big]\,.
\end{align}
The first factor in the second line is antisymmetric in $\ell\leftrightarrow k$, while the second is symmetric.
We can thus use the replacement $Z^{[\ell m]}Z^{[kp]}\rightarrow \frac{1}{2}(Z^{[\ell m]}Z^{[kp]}-Z^{[k m]}Z^{[\ell p]})=\frac{1}{2}Z^{[\ell k]}Z^{[m p]}$
to combine the second with the first line, \emph{i.e.}
\begin{align}
	\hat\cJ_{pm}=  
	\sum_{\stackrel{(\ell,k)\neq m,p}{\text{distinct}}}\Big[&
	3Z^{[\ell k]} Z^{[m p]} (L_{kp}-L_{km})(L_{\ell k}+L_{mp})
	\nonumber\\[-3mm]
	& -3Z^{[\ell m]} Z^{[k p]} \, 
	(L_{m\ell}-L_{km}-L_{\ell p}+L_{kp}) (L_{m\ell}+L_{kp})
	\Big]\,.
\end{align}
Terms which do not depend on $\ell$ and $k$ can be converted to (minus) sums only over the excluded values,
\begin{align}
	\hat\cJ_{pm}= \ &
	3\sum_{\stackrel{(\ell,k)\neq m,p}{\text{distinct}}}\Big[
	Z^{[\ell k]} Z^{[m p]} (L_{kp}-L_{km})L_{\ell k}
	-Z^{[\ell m]} Z^{[k p]} \, 
	\left((L_{kp}-L_{km})L_{m\ell}+(L_{m\ell}-L_{\ell p})L_{kp}\right)\Big]
	\nonumber\\
	&
	-3\sum_{\stackrel{k\neq m,p}{\ell\in\lbrace k,m,p\rbrace}}\!\left[Z^{[\ell k]} Z^{[m p]} (L_{kp}-L_{km})L_{mp}
	-Z^{[\ell m]} Z^{[k p]}(L_{kp}-L_{km})L_{kp}\right]
	\nonumber\\
	&+3\sum_{\stackrel{\ell\neq m,p}{k\in\lbrace \ell,m,p\rbrace}} \!Z^{[\ell m]} Z^{[k p]} (L_{m\ell}-L_{\ell p})L_{m\ell}~.
\end{align}
Upon renaming $\ell\leftrightarrow k$ in the last term of the first line, and in the last line, the lower two lines become 
the missing terms for the summation over $\ell$ in the first line,
\begin{align}
	\hat\cJ_{pm}= \ &
	3\sum_{k\neq m,p}(L_{kp}-L_{km})\sum_\ell\Big[
	Z^{[\ell k]} Z^{[m p]} L_{\ell k}
	-Z^{[\ell m]} Z^{[k p]} \, L_{m\ell}+Z^{[k m]} Z^{[\ell p]}L_{\ell p}\Big]\,.
\end{align}
The sum over $\ell$ is precisely the regularity condition (\ref{eq:constr-4}). 
We conclude that $\hat \cJ_{pm}=0$, thus showing equation (\ref{eq:extr}). Thus, the functional $F(\vec{r})$ is extremized by pole configurations that satisfy the regularity conditions.


\section{Sample solutions}\label{app:sample}

In this appendix, we present a sample of explicit solutions and their free energies. 
To streamline the notation we set $2\pi\alpha^\prime=1$.

Solutions for $L=5, \ldots, 10$ poles, connected by a Higgs branch flow separating complete 5-branes from the brane web, are shown in Table~\ref{tab:one}.
The data for the UV and IR theories are denoted by UV and IR superscripts, respectively, and the IR theories are characterized by the prescription in equation (\ref{eq:HB-IR-res-1}) with $K=2$, (\emph{i.e.} the first two residues are affected, while the remaining ones stay fixed) and where $\xi \in (0,1)$ was chosen arbitrarily.

Solutions with $L=5, \ldots, 10$ poles connected by a Higgs branch flow separating a triple-junction of 5-branes from the brane web are shown in Table~\ref{tab:two}. The IR theories are characterized by the prescription in equation (\ref{eq:HB-IR-res-1}) with $K=3$ (\emph{i.e.} the first three residues are affected, while the remaining ones stay fixed) and $\xi \in (0,1)$ was chosen randomly. The IR theories comprise two sectors for these flows, and both contributions are included in $\Delta F$.

Solutions for UV and IR theories connected by $SU(2)_{R}$-preserving RG flows, for which the UV solutions have $L=3, \ldots, 8$ poles, are shown in Table~\ref{tab:three}.
The IR theories are characterized by the prescription in equation (\ref{eq:mass-IR-res-1}), where we have conveniently chosen $s=1$ and $t=2$  (\emph{i.e.} the first two residues as well as the $(L+1)^{\rm th}$ are affected, while the remaining ones stay fixed). $\alpha,\beta \in (0,1)$ were chosen randomly. 
The IR SCFTs comprise two sectors, one described by a solution with $L+1$ poles and one described by a solution with $3$ poles; both contribute to $\Delta F$.

\begin{turnpage}
\begin{table}
	\renewcommand*{\arraystretch}{0.9}
	{\tiny
		\begin{tabular}{c|c|c|c|c|c}
			$L$ & $\left\{ (p_i,q_i)^{\rm UV} \right\}_{i=1}^{L}$ & $\left\{ r^{\rm UV}_{\ell} \right\}_{\ell=1}^{L}$, $\left\{ r^{\rm IR}_{\ell} \right\}_{\ell=1}^{L}$ & $\xi$ & $F_{\rm UV}$ & $\Delta F/F_{\rm UV}$\\
			\hline
			\hline
			$ 5 $ & $ \left\{
			\begin{array}{ccc}
			(41,-95) ,  & (-41,95) ,  & (96,-2) ,  \\
			(64,66) ,  & (-160,-64)  &  \\
			\end{array}
			\right\} $  & \begin{tabular}{l}$ \left\{
				\begin{array}{ccc}
				1.52073 ,  & -0.657577 ,  & 96.0104 ,  \\
				-1.45615 ,  & 0.0959769  &  \\
				\end{array}
				\right\} $  ~,   $ \left\{
				\begin{array}{ccc}
				1.52073 ,  & -0.657577 ,  & 96.0104 ,  \\
				-0.657577 ,  & 0.384664  &  \\
				\end{array}
				\right\} $  \end{tabular} &  $ 0.661408 $  &  $ -5.96511\times 10^8 \
			$  &  $ -0.868058 $  \\
			\hline
			$ 5 $ & $ \left\{
			\begin{array}{ccc}
			(26,-29) ,  & (-26,29) ,  & (-96,-30) ,  \\
			(65,32) ,  & (31,-2)  &  \\
			\end{array}
			\right\} $  & \begin{tabular}{l}$ \left\{
				\begin{array}{ccc}
				2.23961 ,  & -0.446506 ,  & 0.152611 ,  \\
				-4.29531 ,  & 31.0322  &  \\
				\end{array}
				\right\} $  ~,   $ \left\{
				\begin{array}{ccc}
				2.23961 ,  & -0.446506 ,  & 0.152611 ,  \\
				-4.29531 ,  & 31.0322  &  \\
				\end{array}
				\right\} $  \end{tabular} &  $ 0.349241 $  &  $ -2.21215\times 10^7 \
			$  &  $ -0.519123 $  \\
			\hline
			$ 6 $ & $ \left\{
			\begin{array}{ccc}
			(11,95) ,  & (-11,-95) ,  & (-68,67) ,  \\
			(55,-36) ,  & (-20,-41) ,  & (33,10)  \\
			\end{array}
			\right\} $   & \begin{tabular}{l}  $ \left\{
				\begin{array}{ccc}
				-1.12247 ,  & 0.890892 ,  & -0.409881 ,  \\
				140.31 ,  & 0.678581 ,  & -6.74819  \\
				\end{array}
				\right\} $ , \\  $ \left\{
				\begin{array}{ccc}
				-1.12247 ,  & 0.890892 ,  & -0.409881 ,  \\
				3.35373 ,  & 0.624829 ,  & -6.74819  \\
				\end{array}
				\right\} $  \end{tabular} &   $ 0.77277 $  &  $ -1.6937\times 10^8 $ \
			&  $ -0.811526 $  \\
			\hline
			$ 6 $ & $ \left\{
			\begin{array}{ccc}
			(53,-54) ,  & (-53,54) ,  & (-66,-49) ,  \\
			(-19,38) ,  & (-36,-73) ,  & (121,84)  \\
			\end{array}
			\right\} $   & \begin{tabular}{l}  $ \left\{
				\begin{array}{ccc}
				2.38266 ,  & -0.419699 ,  & 0.330632 ,  \\
				-0.729265 ,  & 0.544358 ,  & -3.19404  \\
				\end{array}
				\right\} $ , \\  $ \left\{
				\begin{array}{ccc}
				2.38266 ,  & -0.419699 ,  & 0.330632 ,  \\
				-0.67691 ,  & 0.526349 ,  & -3.19404  \\
				\end{array}
				\right\} $  \end{tabular} &   $ 0.470538 $  &  $ -3.60159\times 10^8 \
			$  &  $ -0.57412 $  \\
			\hline
			$ 7 $ & $ \left\{
			\begin{array}{cccc}
			(-58,45) ,  & (58,-45) ,  & (-51,-55) ,  & (37,-19) ,  \\
			(-43,-95) ,  & (40,78) ,  & (17,91)  &  \\
			\end{array}
			\right\} $   & \begin{tabular}{l}  $ \left\{
				\begin{array}{cccc}
				-0.34244 ,  & 2.92022 ,  & 0.436485 ,  & 3.96254 ,  \\
				0.912327 ,  & -15.9989 ,  & -1.20411  &  \\
				\end{array}
				\right\} $ , \\  $ \left\{
				\begin{array}{cccc}
				-0.34244 ,  & 2.92022 ,  & 0.436485 ,  & 3.32518 ,  \\
				1.22036 ,  & 28.5766 ,  & -1.29708  &  \\
				\end{array}
				\right\} $  \end{tabular} &   $ 0.805502 $  &  $ -4.56814\times 10^8 \
			$  &  $ -0.793052 $  \\
			\hline
			$ 7 $ & $ \left\{
			\begin{array}{cccc}
			(82,-11) ,  & (-82,11) ,  & (73,-66) ,  & (-56,-42) ,  \\
			(-85,62) ,  & (-97,63) ,  & (165,-17)  &  \\
			\end{array}
			\right\} $   & \begin{tabular}{l}  $ \left\{
				\begin{array}{cccc}
				14.9759 ,  & -0.0667741 ,  & 2.59716 ,  & 0.978604 ,  \\
				-2.56104 ,  & -1.45058 ,  & 19.4631  &  \\
				\end{array}
				\right\} $ , \\  $ \left\{
				\begin{array}{cccc}
				14.9759 ,  & -0.0667741 ,  & 2.59716 ,  & 0.730576 ,  \\
				-0.880799 ,  & -0.706738 ,  & 19.4631  &  \\
				\end{array}
				\right\} $  \end{tabular} &   $ 0.613267 $  &  $ -5.9828\times 10^8 \
			$  &  $ -0.440846 $  \\
			\hline
			$ 8 $ & $ \left\{
			\begin{array}{cccc}
			(5,-29) ,  & (-5,29) ,  & (-73,-30) ,  & (91,23) ,  \\
			(-30,25) ,  & (15,27) ,  & (10,91) ,  & (-13,-136)  \\
			\end{array}
			\right\} $   & \begin{tabular}{l}  $ \left\{
				\begin{array}{cccc}
				1.18717 ,  & -0.842341 ,  & 0.197467 ,  & -124.02 ,  \\
				-0.241073 ,  & -3.3447 ,  & -1.73496 ,  & 0.90897  \\
				\end{array}
				\right\} $ , \\  $ \left\{
				\begin{array}{cccc}
				1.18717 ,  & -0.842341 ,  & 0.197467 ,  & -8.03746 ,  \\
				-0.36205 ,  & -1.69951 ,  & -1.11591 ,  & 0.90897  \\
				\end{array}
				\right\} $  \end{tabular} &   $ 0.0940663 $  &  $ -4.06325\times \
			10^8 $  &  $ -0.0387971 $  \\
			\hline
			$ 8 $ & $ \left\{
			\begin{array}{cccc}
			(20,-67) ,  & (-20,67) ,  & (-99,27) ,  & (16,90) ,  \\
			(-82,56) ,  & (89,1) ,  & (11,7) ,  & (65,-181)  \\
			\end{array}
			\right\} $   & \begin{tabular}{l}  $ \left\{
				\begin{array}{cccc}
				1.34211 ,  & -0.745095 ,  & -0.133918 ,  & -1.19346 ,  \\
				-0.308885 ,  & -178.006 ,  & -3.43406 ,  & 1.42164  \\
				\end{array}
				\right\} $ , \\  $ \left\{
				\begin{array}{cccc}
				1.34211 ,  & -0.745095 ,  & -0.133918 ,  & -1.19346 ,  \\
				-0.308885 ,  & -178.006 ,  & -3.43406 ,  & 1.42164  \\
				\end{array}
				\right\} $  \end{tabular} &   $ 0.272589 $  &  $ -1.18628\times 10^9 \
			$  &  $ -0.171807 $  \\
			\hline
			$ 9 $ & $ \left\{
			\begin{array}{ccccc}
			(-85,44) ,  & (85,-44) ,  & (8,96) ,  & (19,81) ,  & (-99,50) ,  \\
			(-65,100) ,  & (44,29) ,  & (37,33) ,  & (56,-389)  &  \\
			\end{array}
			\right\} $   & \begin{tabular}{l}  $ \left\{
				\begin{array}{ccccc}
				-0.24348 ,  & 4.10712 ,  & -1.0868 ,  & -1.36021 ,  & -0.232393 ,  \
				\\
				-0.541328 ,  & -3.42804 ,  & -66150.1 ,  & 0.539821  &  \\
				\end{array}
				\right\} $ , \\  $ \left\{
				\begin{array}{ccccc}
				-0.24348 ,  & 4.10712 ,  & -1.0868 ,  & -1.41139 ,  & -0.234702 ,  \
				\\
				-0.511857 ,  & -4.45633 ,  & 4.96662\times 10^7 ,  & 0.548581  &  \\
				\end{array}
				\right\} $  \end{tabular} &   $ 0.552663 $  &  $ -6.12649\times 10^9 \
			$  &  $ -0.424702 $  \\
			\hline
			$ 9 $ & $ \left\{
			\begin{array}{ccccc}
			(25,-8) ,  & (-25,8) ,  & (32,-73) ,  & (4,71) ,  & (36,-9) ,  \\
			(-40,45) ,  & (61,90) ,  & (98,28) ,  & (-191,-152)  &  \\
			\end{array}
			\right\} $   & \begin{tabular}{l}  $ \left\{
				\begin{array}{ccccc}
				6.4061 ,  & -0.156101 ,  & 1.53022 ,  & -0.534699 ,  & 9.13953 ,  \\
				-0.285611 ,  & -0.946289 ,  & -19.3576 ,  & 0.349345  &  \\
				\end{array}
				\right\} $ , \\  $ \left\{
				\begin{array}{ccccc}
				6.4061 ,  & -0.156101 ,  & 1.53022 ,  & -1.1237 ,  & 7.59764 ,  \\
				-0.438917 ,  & -2.59386 ,  & 1.57505 ,  & 0.845987  &  \\
				\end{array}
				\right\} $  \end{tabular} &   $ 0.279621 $  &  $ -1.74933\times 10^9 \
			$  &  $ -0.357282 $  \\
			\hline
			$ 10 $ & $ \left\{
			\begin{array}{ccccc}
			(-100,-26) ,  & (100,26) ,  & (9,-5) ,  & (-94,-75) ,  & (-67,-44) , \
			\\
			(64,-57) ,  & (-74,-22) ,  & (28,51) ,  & (-51,94) ,  & (185,58)  \\
			\end{array}
			\right\} $   & \begin{tabular}{l}  $ \left\{
				\begin{array}{ccccc}
				0.127874 ,  & -7.82018 ,  & 3.85913 ,  & 0.926854 ,  & 0.70087 ,  \\
				2.93339 ,  & 0.201932 ,  & -1.2286 ,  & -0.517938 ,  & -6.53239  \\
				\end{array}
				\right\} $ , \\  $ \left\{
				\begin{array}{ccccc}
				0.127874 ,  & -7.82018 ,  & 3.85913 ,  & 0.797353 ,  & 0.586089 ,  \
				\\
				2.80857 ,  & 0.177559 ,  & -1.17688 ,  & -0.438145 ,  & -6.53239  \\
				\end{array}
				\right\} $  \end{tabular} &   $ 0.69155 $  &  $ -2.95868\times 10^9 \
			$  &  $ -0.400017 $  \\
			\hline
			$ 10 $ & $ \left\{
			\begin{array}{ccccc}
			(-95,-17) ,  & (95,17) ,  & (58,-100) ,  & (-69,-72) ,  & (5,24) ,  \\
			(19,44) ,  & (-45,61) ,  & (61,68) ,  & (-96,80) ,  & (67,-105)  \\
			\end{array}
			\right\} $   & \begin{tabular}{l}  $ \left\{
				\begin{array}{ccccc}
				0.0887686 ,  & -11.2652 ,  & 1.73603 ,  & 0.449236 ,  & -1.36238 ,  \\
				-1.66189 ,  & -0.672172 ,  & -2.50651 ,  & -0.372282 ,  & \
				1.82434  \\
				\end{array}
				\right\} $ , \\  $ \left\{
				\begin{array}{ccccc}
				0.0887686 ,  & -11.2652 ,  & 1.73603 ,  & 0.391661 ,  & -1.38343 ,  \\
				-1.72509 ,  & -0.635843 ,  & -2.72983 ,  & -0.309637 ,  & \
				1.82434  \\
				\end{array}
				\right\} $  \end{tabular} &   $ 0.482824 $  &  $ -4.09376\times 10^9 \
			$  &  $ -0.401009 $  \\
		\end{tabular}
		\caption{UV and IR solutions for $L=5, \ldots, 10$ poles connected by a Higgs branch flow separating complete 5-branes from the brane web.\label{tab:one}}}
\end{table}
%
%
%
%
\begin{table}
	\renewcommand*{\arraystretch}{0.9}
	{\tiny
		\begin{tabular}{c|c|c|c|c|c}
			$L$ & $\left\{ (p_i,q_i)^{\rm UV} \right\}_{i=1}^{L}$ & $\left\{ r^{\rm UV}_{\ell} \right\}_{\ell=1}^{L}$, $\left\{ r^{\rm IR}_{\ell} \right\}_{\ell=1}^{L}$ & $\xi$ & $F_{\rm UV}$ & $\Delta F/F_{\rm UV}$\\
			\hline
			\hline
			$ 5 $ & $ \left\{
			\begin{array}{ccc}
			(-25,-104) ,  & (-61,96) ,  & (86,8) ,  \\
			(-13,78) ,  & (13,-78)  &  \\
			\end{array}
			\right\} $   & \begin{tabular}{l} $ \left\{
				\begin{array}{ccc}
				0.788102 ,  & -0.549385 ,  & -21.5464 ,  \\
				-0.910714 ,  & 1.18046  &  \\
				\end{array}
				\right\} $~,  $ \left\{
				\begin{array}{ccc}
				0.788102 ,  & -0.549385 ,  & -21.5464 ,  \\
				-0.910714 ,  & 1.18046  &  \\
				\end{array}
				\right\} $  \end{tabular} &  $ 0.119948 $  &  $ -1.62367\times 10^8 \
			$  &  $ -0.303924 $  \\
			\hline
			$ 5 $ & $ \left\{
			\begin{array}{ccc}
			(42,69) ,  & (53,-30) ,  & (-95,-39) ,  \\
			(-10,23) ,  & (10,-23)  &  \\
			\end{array}
			\right\} $   & \begin{tabular}{l} $ \left\{
				\begin{array}{ccc}
				-1.77938 ,  & 3.79672 ,  & 0.197275 ,  \\
				-0.649176 ,  & 1.52521  &  \\
				\end{array}
				\right\} $~,  $ \left\{
				\begin{array}{ccc}
				-1.77938 ,  & 3.79672 ,  & 0.197275 ,  \\
				-0.649176 ,  & 1.52521  &  \\
				\end{array}
				\right\} $  \end{tabular} &  $ 0.392459 $  &  $ -3.58601\times 10^7 \
			$  &  $ -0.771686 $  \\
			\hline
			$ 6 $ & $ \left\{
			\begin{array}{ccc}
			(-111,14) ,  & (22,-65) ,  & (89,51) ,  \\
			(21,-75) ,  & (44,-22) ,  & (-65,97)  \\
			\end{array}
			\right\} $   & \begin{tabular}{l} $ \left\{
				\begin{array}{ccc}
				-0.0628142 ,  & 1.39419 ,  & -3.75641 ,  \\
				1.13791 ,  & 1.50463\times 10^7 ,  & -1.03109  \\
				\end{array}
				\right\} $   , \\  $ \left\{
				\begin{array}{ccc}
				-0.0628142 ,  & 1.39419 ,  & -3.75641 ,  \\
				1.27701 ,  & 2.12207 ,  & -0.533657  \\
				\end{array}
				\right\} $  \end{tabular} &  $ 0.0792835 $  &  $ -3.93543\times \
			10^8 $  &  $ -0.217314 $  \\
			\hline
			$ 6 $ & $ \left\{
			\begin{array}{ccc}
			(-120,96) ,  & (51,-20) ,  & (69,-76) ,  \\
			(91,-66) ,  & (-36,54) ,  & (-55,12)  \\
			\end{array}
			\right\} $   & \begin{tabular}{l} $ \left\{
				\begin{array}{ccc}
				-0.350781 ,  & 5.28907 ,  & 2.25855 ,  \\
				3.32185 ,  & -0.465255 ,  & -0.107823  \\
				\end{array}
				\right\} $   , \\  $ \left\{
				\begin{array}{ccc}
				-0.350781 ,  & 5.28907 ,  & 2.25855 ,  \\
				3.32163 ,  & -0.46523 ,  & -0.107823  \\
				\end{array}
				\right\} $  \end{tabular} &  $ 0.00426448 $  &  $ -5.73382\times \
			10^7 $  &  $ -0.00823084 $  \\
			\hline
			$ 7 $ & $ \left\{
			\begin{array}{cccc}
			(113,-159) ,  & (-69,66) ,  & (-44,93) ,  & (-27,-11) ,  \\
			(54,-47) ,  & (-19,-10) ,  & (-8,68)  &  \\
			\end{array}
			\right\} $   & \begin{tabular}{l} $ \left\{
				\begin{array}{cccc}
				1.93751 ,  & -0.401257 ,  & -0.633155 ,  & -0.2145 ,  \\
				3.10921\times 10^7 ,  & 1.65736 ,  & -5.32711  &  \\
				\end{array}
				\right\} $   , \\  $ \left\{
				\begin{array}{cccc}
				1.93751 ,  & -0.401257 ,  & -0.633155 ,  & -0.0110367 ,  \\
				234.394 ,  & -0.12988 ,  & -0.88925  &  \\
				\end{array}
				\right\} $  \end{tabular} &  $ 0.623663 $  &  $ -1.61019\times 10^8 \
			$  &  $ -0.755246 $  \\
			\hline
			$ 7 $ & $ \left\{
			\begin{array}{cccc}
			(9,-77) ,  & (5,81) ,  & (-14,-4) ,  & (-46,-32) ,  \\
			(43,7) ,  & (42,44) ,  & (-39,-19)  &  \\
			\end{array}
			\right\} $   & \begin{tabular}{l} $ \left\{
				\begin{array}{cccc}
				1.12369 ,  & -1.06363 ,  & 0.140055 ,  & 0.568313 ,  \\
				10.3541 ,  & 1.95208 ,  & -0.0500019  &  \\
				\end{array}
				\right\} $   , \\  $ \left\{
				\begin{array}{cccc}
				1.12369 ,  & -1.06363 ,  & 0.140055 ,  & 0.465405 ,  \\
				-215800. ,  & -5.644 ,  & 0.230633  &  \\
				\end{array}
				\right\} $  \end{tabular} &  $ 0.201136 $  &  $ -5.99061\times 10^7 \
			$  &  $ -0.00948316 $  \\
			\hline
			$ 8 $ & $ \left\{
			\begin{array}{cccc}
			(23,-86) ,  & (-97,46) ,  & (74,40) ,  & (-89,-28) ,  \\
			(-45,-28) ,  & (76,81) ,  & (16,78) ,  & (42,-103)  \\
			\end{array}
			\right\} $   & \begin{tabular}{l} $ \left\{
				\begin{array}{cccc}
				1.30259 ,  & -0.225099 ,  & -3.95297 ,  & 0.13771 ,  \\
				0.310405 ,  & -1.67626 ,  & -0.639351 ,  & 1.48771  \\
				\end{array}
				\right\} $   , \\  $ \left\{
				\begin{array}{cccc}
				1.30259 ,  & -0.225099 ,  & -3.95297 ,  & 0.137161 ,  \\
				0.310076 ,  & -1.67599 ,  & -0.638493 ,  & 1.48771  \\
				\end{array}
				\right\} $  \end{tabular} &  $ 0.00678719 $  &  $ -1.61708\times \
			10^9 $  &  $ -0.0107519 $  \\
			\hline
			$ 8 $ & $ \left\{
			\begin{array}{cccc}
			(-55,-24) ,  & (-12,96) ,  & (67,-72) ,  & (-32,23) ,  \\
			(-15,-83) ,  & (-78,28) ,  & (63,-66) ,  & (62,98)  \\
			\end{array}
			\right\} $   & \begin{tabular}{l} $ \left\{
				\begin{array}{cccc}
				0.208681 ,  & -0.882782 ,  & 2.29655 ,  & -0.254467 ,  \\
				0.547375 ,  & -0.0828985 ,  & 0.908948 ,  & -1.81597  \\
				\end{array}
				\right\} $   , \\  $ \left\{
				\begin{array}{cccc}
				0.208681 ,  & -0.882782 ,  & 2.29655 ,  & -0.264643 ,  \\
				0.546107 ,  & -0.0826469 ,  & 0.935056 ,  & -1.81597  \\
				\end{array}
				\right\} $  \end{tabular} &  $ 0.2303 $  &  $ -1.09057\times 10^9 $ \
			&  $ -0.300849 $  \\
			\hline
			$ 9 $ & $ \left\{
			\begin{array}{ccccc}
			(0,7) ,  & (-27,45) ,  & (27,-52) ,  & (31,91) ,  & (-90,-36) ,  \\
			(24,78) ,  & (-38,23) ,  & (-30,-41) ,  & (103,-115)  &  \\
			\end{array}
			\right\} $   & \begin{tabular}{l} $ \left\{
				\begin{array}{ccccc}
				-1. ,  & -0.56619 ,  & 1.646 ,  & -1.90557 ,  & 0.10587 ,  \\
				-1.674 ,  & -0.294669 ,  & 1.02685 ,  & 2.23811  &  \\
				\end{array}
				\right\} $   , \\  $ \left\{
				\begin{array}{ccccc}
				-1. ,  & -0.56619 ,  & 1.646 ,  & -1.99782 ,  & 0.134073 ,  \\
				-1.73189 ,  & -0.292762 ,  & 1.01926 ,  & 2.23811  &  \\
				\end{array}
				\right\} $  \end{tabular} &  $ 0.218455 $  &  $ -8.41213\times 10^8 \
			$  &  $ -0.144715 $  \\
			\hline
			$ 9 $ & $ \left\{
			\begin{array}{ccccc}
			(141,22) ,  & (-94,35) ,  & (-47,-57) ,  & (-67,-58) ,  & (-59,94) , \
			\\
			(55,71) ,  & (36,93) ,  & (-97,-94) ,  & (132,-106)  &  \\
			\end{array}
			\right\} $   & \begin{tabular}{l} $ \left\{
				\begin{array}{ccccc}
				-12.8957 ,  & -0.18013 ,  & 0.471549 ,  & 0.33791 ,  & -0.382773 ,  \\
				-1.27594 ,  & -0.846321 ,  & -0.0301658 ,  & 2.84238  &  \\
				\end{array}
				\right\} $   , \\  $ \left\{
				\begin{array}{ccccc}
				-12.8957 ,  & -0.18013 ,  & 0.471549 ,  & 0.329389 ,  & -0.377873 ,  \
				\\
				-1.4478 ,  & -0.905088 ,  & 0.0014699 ,  & 2.84238  &  \\
				\end{array}
				\right\} $  \end{tabular} &  $ 0.464149 $  &  $ -5.83744\times 10^9 \
			$  &  $ -0.470571 $  \\
			\hline
			$ 10 $ & $ \left\{
			\begin{array}{ccccc}
			(-51,102) ,  & (67,-55) ,  & (-16,-47) ,  & (61,-86) ,  & (-48,60) , \
			\\
			(37,28) ,  & (-61,-36) ,  & (-86,-17) ,  & (49,-49) ,  & \
			(48,100)  \\
			\end{array}
			\right\} $   & \begin{tabular}{l} $ \left\{
				\begin{array}{ccccc}
				-0.618034 ,  & 2.79424 ,  & 0.715931 ,  & 1.76614 ,  & -0.293336 ,  \\
				-2.58274\times 10^7 ,  & 0.476415 ,  & 0.302445 ,  & 0.824145 ,  & \
				-1.69735  \\
				\end{array}
				\right\} $   , \\  $ \left\{
				\begin{array}{ccccc}
				-0.618034 ,  & 2.79424 ,  & 0.715931 ,  & 1.44384 ,  & -0.434273 ,  \\
				-9.99286 ,  & 0.403255 ,  & 0.152776 ,  & 1.60986 ,  & -1.58923  \\
				\end{array}
				\right\} $  \end{tabular} &  $ 0.948093 $  &  $ -2.32359\times 10^9 \
			$  &  $ -0.722034 $  \\
			\hline
			$ 10 $ & $ \left\{
			\begin{array}{ccccc}
			(-110,102) ,  & (14,-44) ,  & (96,-58) ,  & (13,-97) ,  & (20,-89) , \
			\\
			(12,-79) ,  & (-2,-92) ,  & (-53,-91) ,  & (91,-94) ,  & (-81,542)  \\
			\end{array}
			\right\} $   & \begin{tabular}{l} $ \left\{
				\begin{array}{ccccc}
				-0.392288 ,  & 1.36758 ,  & 3.58898 ,  & 0.807556 ,  & 1.09511 ,  \\
				0.868635 ,  & 0.501708 ,  & 0.12168 ,  & 4.70283\times 10^6 ,  & \
				-2.58745  \\
				\end{array}
				\right\} $   , \\  $ \left\{
				\begin{array}{ccccc}
				-0.392288 ,  & 1.36758 ,  & 3.58898 ,  & 0.655044 ,  & 1.03151 ,  \\
				0.735122 ,  & 0.300381 ,  & -0.0746285 ,  & 6.95946\times 10^6 ,  & \
				-1.78367  \\
				\end{array}
				\right\} $  \end{tabular} &  $ 0.739462 $  &  $ -9.69288\times 10^9 \
			$  &  $ -0.683226 $  
		\end{tabular}
		\caption{UV, IR solutions with $L=5, \ldots, 10$ poles connected by a Higgs branch flow separating triple-junctions from the brane web.\label{tab:two}}}
\end{table}
%
%
%
%
\begin{table}
	\renewcommand*{\arraystretch}{0.9}
	{\tiny
		\begin{tabular}{c|c|c|c|c|c|c}
			$L$ & $\left\{ (p_i,q_i)^{\rm UV} \right\}_{i=1}^{L}$ & $\left\{ r^{\rm UV}_{\ell} \right\}_{\ell=1}^{L}$, $\left\{ r^{\rm IR}_{\ell} \right\}_{\ell=1}^{L+1}$ & $\alpha$ &$\beta$ & $F_{\rm UV}$ & $\Delta F/F_{\rm UV}$\\
			\hline
			\hline
			$ 3 $ & $ \left\{
			\begin{array}{ccc}
			(49,69),  & (36,-21), & (-85,-48)  \\
			\end{array}
			\right\} $  & 
			\begin{tabular}{l}  
				$ \left\{
				\begin{array}{ccc}
				-1,  & 0,  &				1 \\
				\end{array}
				\right\} $ , $ \left\{
				\begin{array}{cccc}
				-1.93665 ,  & 3.69892 , &
				0.262846 ,  & -8.47455  \\
				\end{array}
				\right\} $  \end{tabular} &  $ 0.76 $  &  $ 0.59 $  &  $ -5.07289\times 10^6 $  &  $ 0.527495 $  \\
			\hline
			$ 3 $ & $ \left\{
			\begin{array}{ccc}
			(-76,-73) ,  & (64,-84) , &	(12,157)  \\
			\end{array}
			\right\} $  & \begin{tabular}{l}  $ \left\{
				\begin{array}{ccc}
				-1 ,  & 0 ,  &	1  \\
				\end{array}
				\right\} $ , $ \left\{
				\begin{array}{cccc}
				0.402469 ,  & 2.01908 , &-1.07935 ,  & 1.23992  \\
				\end{array}
				\right\} $  \end{tabular} &  $ 0.05 $  &  $ 0.16 $  &  $ -5.02453\times 10^7 $  &  $ 0.0488212 $  \\
			\hline
			$ 4 $ & $ \left\{
			\begin{array}{cc}
			(-40,-75) ,  & (-56,136) ,  \\
			(77,-71) ,  & (19,10)  \\
			\end{array}
			\right\} $  & \begin{tabular}{l}  $ \left\{
				\begin{array}{cc}
				0.6 ,  & -0.31124 ,  \\
				2.55969 ,  & -4.04709  \\
				\end{array}
				\right\} $   , \\  $ \left\{
				\begin{array}{ccc}
				0.6 ,  & -2.05184 ,  & 2.55969 ,  \\
				-4.04709 ,  & -0.407393  &  \\
				\end{array}
				\right\} $  \end{tabular} &  $ 0.98 $  &  $ 0.93 $  &  $ -5.20266\times 10^7 $  &  $ 0.325837 $  \\
			\hline
			$ 4 $ & $ \left\{
			\begin{array}{cc}
			(-2,-36) ,  & (-6,-94) ,  \\
			(93,58) ,  & (-85,72)  \\
			\end{array}
			\right\} $  & \begin{tabular}{l}  $ \left\{
				\begin{array}{cc}
				0.945986 ,  & 0.875789 ,  \\
				-3.49317 ,  & -0.366607  \\
				\end{array}
				\right\} $   , \\  $ \left\{
				\begin{array}{ccc}
				0.945986 ,  & 0.869831 ,  & -3.49317 ,  \\
				-0.366607 ,  & 0.88699  &  \\
				\end{array}
				\right\} $  \end{tabular} &  $ 0.52 $  &  $ 0.84 $  &  $ -5.56051\times 10^7 $  &  $ 0.000431119 $  \\
			\hline
			$ 5 $ & $ \left\{
			\begin{array}{ccc}
			(45,-6) ,  & (-126,7) ,  & (41,-49) ,  \\
			(-9,30) ,  & (49,18)  &  \\
			\end{array}
			\right\} $  & \begin{tabular}{l}  $ \left\{
				\begin{array}{ccc}
				6.37105 ,  & 0.704959 ,  & 2.14062 ,  \\
				-0.744031 ,  & -5.62231  &  \\
				\end{array}
				\right\} $   , \\  $ \left\{
				\begin{array}{ccc}
				7.22446 ,  & 0.548949 ,  & 2.14062 ,  \\
				-0.744031 ,  & -5.62231 ,  & 0.64397  \\
				\end{array}
				\right\} $  \end{tabular} &  $ 0.41 $  &  $ 0.74 $  &  $ -3.39657\times 10^7 $  &  $ 0.31831 $  \\
			\hline
			$ 5 $ & $ \left\{
			\begin{array}{ccc}
			(88,-49) ,  & (-61,24) ,  & (-30,76) ,  \\
			(-24,1) ,  & (27,-52)  &  \\
			\end{array}
			\right\} $  & \begin{tabular}{l}  $ \left\{
				\begin{array}{ccc}
				3.85148 ,  & -0.164132 ,  & -0.680353 ,  \\
				-0.0208243 ,  & 0.336022  &  \\
				\end{array}
				\right\} $   , \\  $ \left\{
				\begin{array}{ccc}
				3.85148 ,  & -0.172337 ,  & -0.680353 ,  \\
				-0.0208243 ,  & 0.370241 ,  & -0.13266  \\
				\end{array}
				\right\} $  \end{tabular} &  $ 0.16 $  &  $ 0.78 $  &  $ -3.12429\times 10^7 $  &  $ 0.291831 $  \\
			\hline
			$ 6 $ & $ \left\{
			\begin{array}{ccc}
			(47,-76) ,  & (-96,-49) ,  & (-11,-69) ,  \\
			(-29,44) ,  & (32,81) ,  & (57,69)  \\
			\end{array}
			\right\} $  & \begin{tabular}{l}  $ \left\{
				\begin{array}{ccc}
				4.06201 ,  & -0.0301757 ,  & 0.853207 ,  \\
				-0.538574 ,  & -1.16889 ,  & -2.12317  \\
				\end{array}
				\right\} $   , \\  $ \left\{
				\begin{array}{cccc}
				17.7224 ,  & -0.0755114 ,  & 0.853207 ,  & -0.538574 ,  \\
				-1.16847 ,  & -2.12317 ,  & 3.31567  &  \\
				\end{array}
				\right\} $  \end{tabular} &  $ 0.91 $  &  $ 0.23 $  &  $ -3.03209\times 10^8 $  &  $ 0.290211 $  \\
			\hline
			$ 6 $ & $ \left\{
			\begin{array}{ccc}
			(30,-68) ,  & (47,166) ,  & (25,61) ,  \\
			(-82,-51) ,  & (-96,-64) ,  & (76,-44)  \\
			\end{array}
			\right\} $  & \begin{tabular}{l}  $ \left\{
				\begin{array}{ccc}
				1.53417 ,  & -0.816438 ,  & -1.49056 ,  \\
				0.308714 ,  & 0.386039 ,  & 3.72314  \\
				\end{array}
				\right\} $   , \\  $ \left\{
				\begin{array}{cccc}
				1.53417 ,  & -0.781773 ,  & -1.49056 ,  & 0.25905 ,  \\
				0.34285 ,  & 3.72314 ,  & -0.975598  &  \\
				\end{array}
				\right\} $  \end{tabular} &  $ 0.1 $  &  $ 0.78 $  &  $ -5.15112\times 10^8 $  &  $ 0.108757 $  \\
			\hline
			$ 7 $ & $ \left\{
			\begin{array}{cccc}
			(-219,35) ,  & (-9,55) ,  & (85,-38) ,  & (42,9) ,  \\
			(34,-56) ,  & (-8,93) ,  & (75,-98)  &  \\
			\end{array}
			\right\} $  & \begin{tabular}{l}  $ \left\{
				\begin{array}{cccc}
				0.765802 ,  & -0.614318 ,  & 4.68704 ,  & -17.7392 ,  \\
				1.72934 ,  & -0.917672 ,  & 2.02455  &  \\
				\end{array}
				\right\} $   , \\  $ \left\{
				\begin{array}{cccc}
				0.994092 ,  & -0.648232 ,  & 4.68704 ,  & -14.7329 ,  \\
				1.73781 ,  & -0.917672 ,  & 2.02455 ,  & 0.722395  \\
				\end{array}
				\right\} $  \end{tabular} &  $ 0.78 $  &  $ 0.61 $  &  $ -9.22145\times 10^8 $  &  $ 0.21308 $  \\
			\hline
			$ 7 $ & $ \left\{
			\begin{array}{cccc}
			(68,88) ,  & (-2,-48) ,  & (-74,89) ,  & (95,22) ,  \\
			(-82,-76) ,  & (71,93) ,  & (-76,-168)  &  \\
			\end{array}
			\right\} $  & \begin{tabular}{l}  $ \left\{
				\begin{array}{cccc}
				-2.09447 ,  & 5.86503 ,  & -0.46905 ,  & -8.75064 ,  \\
				0.543158 ,  & -2.02155 ,  & 2.03823  &  \\
				\end{array}
				\right\} $   , \\  $ \left\{
				\begin{array}{cccc}
				-2.06537 ,  & 9.46876 ,  & -0.46905 ,  & -8.75064 ,  \\
				0.9581 ,  & -2.02155 ,  & 3.36475 ,  & -4.19094  \\
				\end{array}
				\right\} $  \end{tabular} &  $ 0.84 $  &  $ 0.88 $  &  $ -1.39638\times 10^9 $  &  $ 0.283667 $  \\
			\hline
			$ 8 $ & $ \left\{
			\begin{array}{cccc}
			(-220,183) ,  & (69,25) ,  & (-8,52) ,  & (-55,-3) ,  \\
			(39,-78) ,  & (60,-95) ,  & (43,-44) ,  & (72,-40)  \\
			\end{array}
			\right\} $  & \begin{tabular}{l}  $ \left\{
				\begin{array}{cccc}
				-0.279993 ,  & -2.06941 ,  & -0.857919 ,  & 0.0272525 ,  \\
				0.340598 ,  & 0.495622 ,  & 1.03228 ,  & 3.85913  \\
				\end{array}
				\right\} $   , \\  $ \left\{
				\begin{array}{ccccc}
				-0.22106 ,  & -2.09015 ,  & -0.857919 ,  & 0.0272525 ,  & 0.326212 , \
				\\
				0.478023 ,  & 1.00764 ,  & 3.85913 ,  & -0.277575  &  \\
				\end{array}
				\right\} $  \end{tabular} &  $ 0.98 $  &  $ 0.12 $  &  $ -1.17444\times 10^9 $  &  $ 0.0998954 $  \\
			\hline
			$ 8 $ & $ \left\{
			\begin{array}{cccc}
			(-43,-52) ,  & (22,-94) ,  & (-73,-78) ,  & (-99,27) ,  \\
			(-7,53) ,  & (30,-93) ,  & (-10,18) ,  & (180,219)  \\
			\end{array}
			\right\} $  & \begin{tabular}{l}  $ \left\{
				\begin{array}{cccc}
				0.515074 ,  & 1.29848 ,  & 0.433739 ,  & -0.133918 ,  \\
				-0.876609 ,  & 1.4857 ,  & -0.555227 ,  & -12.4806  \\
				\end{array}
				\right\} $   , \\  $ \left\{
				\begin{array}{ccccc}
				0.503936 ,  & 1.49054 ,  & 0.433739 ,  & -0.133918 ,  & -0.876609 ,  \
				\\
				1.69323 ,  & -0.55893 ,  & -9.21582 ,  & 0.775377  &  \\
				\end{array}
				\right\} $  \end{tabular} &  $ 0.9 $  &  $ 0.29 $  &  $ -2.13494\times 10^9 $  &  $ 0.0532703 $  \\
		\end{tabular}
		\caption{UV, IR solutions connected by $SU(2)_{R}$-preserving RG flows, whose UV theories have $L=3, \ldots, 8$ poles.\label{tab:three}}}
\end{table}
\end{turnpage}

\clearpage

\bibliography{notes}
\end{document}